\documentclass[final,10pt,3p,times]{elsarticle}
\journal{Computer Networks}

\usepackage{amssymb}

\usepackage{amsmath}
\usepackage{algorithmicx}
\usepackage{lipsum}
\usepackage{times}
\usepackage{url}
\usepackage{multirow}
\usepackage{algpseudocode}
\usepackage{algorithm}
\usepackage{rotating}
\usepackage{subcaption} 
\usepackage{epstopdf}
\usepackage{commath}
\allowdisplaybreaks

\usepackage[table]{xcolor}
\usepackage{color, colortbl}
\definecolor{Gray}{gray}{0.9}

\newcommand{\coldscr}{\cellcolor{Gray}}
\definecolor{LightCyan}{rgb}{0.88,1,1}
\algnewcommand\algorithmicinput{\textbf{INPUT:}}
\algnewcommand\INPUT{\item[\algorithmicinput]}
\algnewcommand\algorithmicoutput{\textbf{OUTPUT:}}
\algnewcommand\OUTPUT{\item[\algorithmicoutput]}

\algdef{SE}[DOWHILE]{Do}{doWhile}{\algorithmicdo}[1]{\algorithmicwhile\ #1}

\makeatletter
\newcounter{algorithmbis}
\renewcommand{\thealgorithmbis}{\arabic{algorithmbis}}
\def\algorithmbis{\@ifnextchar[{\@algorithmbisa}{\@algorithmbisb}}
\def\@algorithmbisa[#1]{%
  \refstepcounter{algorithmbis}
  \trivlist
  \leftmargin\z@
  \itemindent\z@
  \labelsep\z@
  \item[\parbox{0.49\textwidth}{%
    \hrule
    \noindent\strut\textbf{Algorithm \thealgorithmbis} #1
    \hrule
  }]\hfil\vskip+0em%
}
\def\@algorithmbisb{\@algorithmbisa[]}

\makeatother

\begin{document}
	\begin{frontmatter}

\title{Joint Failure Recovery, Fault Prevention, and Energy-efficient Resource Management for Real-time SFC in Fog-supported SDN}

\author[label1]{Mohammad M. Tajiki}
\address[label1]{Department of Electronic Engineering, University of Rome – Tor Vergata, Via del Politecnico 1, 00133, Rome, Italy}
\ead{mhmtjk01@uniroma2.it}
\author[label2]{Mohammad Shojafar}
\address[label2]{Department of Mathematics,
University of Padua, Via Trieste 63, 35131, Padua, Italy}
\ead{mohammad.shojafar@math.unipd.it}
\author[label4]{Behzad Akbari}
\address[label4]{Deparment of Electrical and Computer Engineering, Tarbiat Modares University, Tehran, Iran}
\ead{b.akbari@modares.ac.ir}
\author[label1]{Stefano Salsano}
\ead{stefano.salsano@uniroma2.it}
\author[label2]{Mauro Conti}
\ead{conti@math.unipd.it}
\author[label5]{Mukesh Singhal}
\address[label5]{Department of Electrical and Computer Engineering, University of California- Merced, Merced, CA 95343, USA}
\ead{msinghal@ucmerced.edu}

\begin{abstract}
Middleboxes have become a vital part of modern networks by providing services such as load balancing, optimization of network traffic, and content filtering. A sequence of middleboxes comprising a logical service is called a \textit{Service Function Chain (SFC)}. In this context, the main issues are to maintain an acceptable level of network path survivability and a fair allocation of the resource between different demands in the event of faults or failures. In this paper, we focus on the problems of traffic engineering, failure recovery, fault prevention, and SFC with reliability and energy consumption constraints in Software Defined Networks (SDN). These types of deployments use Fog computing as an emerging paradigm to manage the distributed small-size traffic flows passing through the SDN-enabled switches (possibly Fog Nodes). The main aim of this integration is to support service delivery in real-time, failure recovery, and fault-awareness in an SFC context. Firstly, we present an architecture for Failure Recovery and Fault Prevention called FRFP; this is a multi-tier structure in which the real-time traffic flows pass through SDN-enabled switches to jointly decrease the network side-effects of flow rerouting and energy consumption of the Fog Nodes. We then mathematically formulate an optimization problem called the Optimal Fog-Supported Energy-Aware SFC rerouting algorithm (OFES) and propose a near-optimal heuristic called Heuristic OFES (HFES) to solve the corresponding problem in polynomial time. In this way, the energy consumption and the reliability of the selected paths are optimized, while the Quality of Service (QoS) constraints are met and the network congestion is minimized. In a reliability context, the focus of this work is on fault prevention; however, since we use a reallocation technique, the proposed scheme can be used as a failure recovery scheme. We compare the performance of HFES and OFES in terms of power consumption, average path length, fault probability, network side-effects, link utilization, and Fog Node utilization. Additionally, we analyze the computational complexity of HFES. We use a real-world network topology to evaluate our algorithm. The simulation results show that the heuristic algorithm is applicable to large-scale networks.
\end{abstract}
\end{frontmatter}

\section{Introduction}\label{sec:introduction}

   Network Function Virtualization (NFV) technology plays an important role in industrial traffic management through a chain of different hardware services running on the middle-boxes (e.g., IDS, proxy, deep packet inspection, and firewall), which is called Service Function Chaining (SFC). In particular, NFV replaces hardware middle-boxes with flexible and innovative software applications known as Virtual Network Functions (VNFs) to reduce the CAPEX/OPEX costs, optimize network operations, and increase the service usage elasticity~\cite{fischer2013virtual}. Also, NFV can decrease the dependence on expensive network equipment vendor solutions, by replacing network functions with software implementations running on low-cost multi-purpose hardware~\cite{naudts2016deploying}.
   Focusing on SFC, a chain of economical VNFs provides the same packet processing functions at the desired throughput~\cite{ghaznavi2017distributed}. Besides, chains are required to process large volumes of traffic within a very short period of time to facilitate real-time streaming applications that comprise the majority of traffic in today's networks. SFCs require attention to avoid cascading threats as well as controller protections, especially for the software applications across the SDN switches that are corporate with the server virtualization and virtual machines (VMs)~\cite{amoroso2017software}. Failure to provide the desired throughput of an SFC may lead to violating the service level agreements (SLAs), incurring high penalties. Hence, achieving the high throughput of ordered VNFs (i.e., it can be interpreted as SFC) is of paramount importance.
    
    In SDN-based SFC, controllers are expected to provide high availability for the traffic flows. Hence, they are hungry to use innovative solutions (possibly in routing and rerouting) to preserve the availability of the chains from failure. On the other hand, the Fog computing paradigm is defined to deploy computing resources closer to end users. Fog computing at the edge can rapidly compute and organize small instance processes locally and move relevant on-demand processing data flow from the incident geographical location to core platforms such as Amazon Web Services \cite{gargees2017incident}. Moreover, some SDN-enabled switches that are located geographically near to the users are played edge switches (nodes or Fog Nodes) to address small-size flows with limited response time SLAs and deliver high user Quality of Experiences (QoEs) like ~\cite{peng2016fog} and~\cite{liang2017integrated}. As a consequence, Fog Nodes are connected with virtualized SDN-enabled switches, which run atop servers or data centers at the edge of the access network. These switches can easily handle such flows within the low latency. The big issue behind this technology is how to control the Fog Node when faced with failure or side-effects of faults? Therefore, presenting failure recovery and fault-aware solutions in fog-supported SDN/NFV-based SFC is an essential phenomenon that must be addressed. Recently, several technical/practical works have been presented in the literature in order to address such limitations in SDNs/NFVs. Most of these works address the fault-handling process, from fault detection and prevention to failure recovery, but to the best of our knowledge none of them jointly addresses the SFC QoSs fault-recovery, fault minimization targeting prevention, path reliability, and energy minimization over SDN-enabled switches, all of which our solution covers. To cope with the problem, several questions arise: Is it possible to propose fault tolerance routing and rerouting algorithms for real-time guarantee time-triggered traffic by supporting QoS SFC and minimize the energy in SDNs/NFVs? How to guarantee the elasticity of such solutions, which can be applied, in real scenarios? Can we assure that the presented algorithm could swiftly update itself for the dynamically changing time-triggered traffic? 
    
    \subsection{Contributions}\label{sez:1.1}
   Motivated by the aforementioned considerations, we address the problem of SFC using the NFV concept in fog-supported SDNs with reliability, QoS, and energy consumption considerations. To this end, we proposed a routing architecture based on the SDN concept with a focus on the failure aspect of network devices. We mathematically formulate the problem of traffic engineering when the network devices have a variable fault probability during the time slots and use an Integer Linear Programming (ILP) solver to optimally solve the corresponding optimization problem. Thereafter, we propose an efficient heuristic algorithm to handle the scalability issue over large-scale networks. Our main contributions are summarized as follows:
    \begin{itemize}
        \item We propose a new fault-aware routing architecture for SFC problems with energy consideration. The architecture is proposed for SDN networks and supports fog nodes. 
        We mathematically formulate the failure recovery and fault minimization problem. The corresponding problem is in the form of ILP. We consider the impact of each flow on other flows to surpass the resource fragmentation in networks with big-size flows.
        \item Our proposed scheme optimizes the probability of failure in the networking devices (switches) and, in the event of failure in a Fog Node(s) and/or switch(es), reconfigures the network in a real-time manner.
        \item We propose a suboptimal heuristic to solve the scalability issues of the aforementioned optimization problem. The proposed solution is an adaptive approach that is applicable to real-world networks.
        \item In order to evaluate the proposed algorithm, we exploit a real-world network topology. Additionally, we test the performance of the proposed algorithm over different traffic patterns using a demand generator. To this end, the impacts of the flow size, number of Fog Nodes, and length of the required VNF on the performance of the proposed algorithm are evaluated.
    \end{itemize}
    
    \subsection{Organization}
    The remainder of this paper is organized as follows. Section~\ref{sec:relatedwork} presents a holistic literature. Section~\ref{sec:refscenario} presents the problem definition, related assumptions and overviews the considered architecture and its main components while Section~\ref{sec:model} presents the system model. The problem of jointly managing the energy consumption and the network side-effect of rerouting flows triggered by fatigue processes is formulated as an ILP in Section~\ref{sec:formulation}. Section~\ref{sec:HFNR} details the proposed heuristic algorithm, HNFR, and its computational complexity. The considered scenarios and the setting of the input parameters are detailed in Section~\ref{sec:scenariodescription}. The obtained results are detailed in Section~\ref{sec:results}. Finally, Section~\ref{sec:conclusions} concludes the paper with some final remarks and outlines open research problems.
    
\section{Related Work}\label{sec:relatedwork}
In the following, we briefly discuss the main literature on NFVs/SDNs SFC related to our work. We first describe solutions targeting the SFC. Then, we move our attention to research works targeting the management of failures and faults for various injected traffic flows in SDNs/NFVs. Finally, we investigate the energy-aware fog-supported solutions in SDNs/NFVs.
\subsection{SFC solutions in SDNs/NFVs}\label{sec:relatedwork.1}
Consequently, numerous works focus on providing SFC in SDNs. An SFC taxonomy that considers architecture and performance dimensions as the basis for the subsequent state-of-the-art analysis is introduced in \cite{medhat2017service}. 

The authors of \cite{bhamare2017optimal} study the problem of deploying SFCs over NFV architecture. Specifically, they investigate the VNF placement problem for the optimal SFC formation across geographically distributed clouds. Moreover, they set up the problem of minimizing inter-cloud traffic and response time in a multi-cloud scenario as an ILP optimization problem, along with some other constraints such as total deployment costs and SLAs.

Moreover, in \cite{reddy2016robust} an optimization model based on the concept of $\Gamma$-robustness is proposed. They focus on dealing with the uncertainty of the traffic demand. The authors of \cite{zhang2016co} propose a heuristic algorithm to find a solution for service chaining. It employs two-step flow selection when an SFC with multiple network functions needs to scale out. Furthermore, the authors in \cite{abdelsalam2017implementation} introduce a VNF chaining which is implemented through segment routing in a Linux-based infrastructure. To this end, they exploit an IPv6 Segment Routing (SRv6) network programming model to support SFC in an NFV scenario. The authors of \cite{kulkarni2017neo} propose a scheme which provides flexibility, ease of configuration and adaptability to relocate the service functions with a minimal control plane overhead.
    
Besides, the authors of \cite{bari2015orchestrating} use ILP to determine the required number and placement of VNFs that optimize network CAPEX/OPEX costs without violating SLAs. In \cite{even2016approximation} an approximation algorithm for path computation and function placement in SDNs is proposed. Similar to \cite{bari2015orchestrating}, they proposed a randomized approximation algorithm for path computation and function placement. In \cite{ghaznavi2016service} an optimization model to deploy a chain in a distributed manner is developed. Their proposed model abstracts heterogeneity of VNF instances and allows them to deploy a chain with custom throughput without worrying about individual VNFs’ throughput. The paper \cite{rost2016service} considers the offline batch embedding of multiple service chains. They consider the objectives of maximizing the profit by embedding an optimal subset of requests or minimizing the costs when all requests need to be embedded. Reference \cite{jiang2012joint} solves a joint route selection and VM placement problem. They design an offline algorithm to solve a static VM placement problem and an online solution traffic routing. They expand the technique of Markov approximation to achieve their objectives. 

Recently, the authors in \cite{TII2017SDN} presented a scheduling and routing solution in SDN/NFV time-triggered flows. In detail, they approximate the optimal solution over a corresponding static scheduling problem and solve it using ILP. As in our approach, hard constraints on the overall execution times are considered by \cite{TII2017SDN}. However, we point out that, unlike our approach: (i) the focus of \cite{TII2017SDN} is on the traffic routing and scheduling between SDN-enabled switches per time-flow, so that the resulting flow scheduler does not support, by design, failure and fault tolerance per link and switch of data time-flow; (ii) the joint flow and computing rate mapping afforded in \cite{TII2017SDN} is, by design, static; (iii) the scheduler in \cite{TII2017SDN} does not perform real-time reconfiguration rerouting, real-time traffic hosted by the serving controller; (iv) the work in \cite{TII2017SDN} does not consider SFC and rerouting; and (v) the scheduler in \cite{TII2017SDN} does not enforce per-flow QoS guarantees on the limited time minimum energy and/or the minimum side-effect. Although the aforementioned solutions are interesting, however, none of them considers the problem of service chaining with respect to the energy consumption of the VMs.

\subsection{Failure recovery and fault-aware solutions in SDNs/NFVs}\label{sec:relatedwork.2}

The available literature ranges from the joint problem of fault-aware distributing and routing the traffic flows/content in SDNs/NFVs infrastructure \cite{sterbenz2010resilience,kreutz2013towards} to the problem of fault detection and recovery solutions in SDNs/NFVs~\cite{vilchez2014self,fonseca2017survey}. 
In detail, in \cite{kreutz2013towards} the authors analyze the fault tolerance over SDN. They present a discussion about fault tolerance and failure happening in the OpenFlow (OF) protocol that is applied in SDNs. Specifically, they propose a link/node failure detection and failure recovery method in the data plane that can be controlled through the controller. However, they do not present any discussion about the application plane side-effect and do not cover the SFC fault-awareness. 

In \cite{sharma2013openflow}, they present a controller-based fault recovery solution that covers path-failure detection and preconfigured backup paths. However, we point out that, unlike our approach: (i) the focus of \cite{sharma2013openflow} is on presenting the network configuration in order to manage the traffic flows, which is not an effective solution, by design, in real scenarios;  and (ii) the presented fault prevention method in \cite{sharma2013openflow} does not support the SFC over the SDN-enabled switches. The authors in \cite{van2014fast} propose a solution to quickly detect link failures in order to increase the fault tolerance by combining the flow retrieval which is achieved through analyzing the protection switching times and using a fast protection method. Interestingly, this paper supports the fault minimization over the links and addresses the end-to-end fault tolerance method per flow, but not radically. Overall, the contribution does not afford, by design, jointly the QoSs in the node and link of SDN and does not support the SFC fault minimization, both of which are adopted in this paper.

Besides, authors in \cite{turchetti2015implementation} present NFV-FD, a fault-tolerant unreliable failure detector that is adapted based on information (it includes communication links states and the flow characteristics) obtained from an SDN. The paper presents flavor of novelties, but it fails to address the SFC traffic flows. Moreover, our solution utilizes a network equipment fault-aware technique that spreads out the fault tolerance process all over the components running in the SDN. In \cite{reitblatt2013fattire}, authors applied novel rule-based programming language presented in \cite{guha2013machine} to talk between the controller and the data plane to manage the adopted in-network fast-fail over mechanisms of incoming traffic flows in FatTire programs. Although this method is an interesting step toward to the fault-aware SDN traffic flow policy management, it suffers from fault recovery and fault prevention that matter in our solution.

\subsection{Energy-aware Fog-supported solutions in SDNs/NFVs}\label{sec:relatedwork3}
Numerous works address the switch energy efficiency and energy-aware routing strategies in SDNs/NFVs \cite{heller2010elastictree,bolla2014dropv2,etsi2014network,zhu2016joint,awad2017energy}. In detail, the authors in \cite{heller2010elastictree} present a network-wide energy-aware routing method using OF maximizing aggregate network utilization and optimized load balancing in SDN. Their practical solution has problem with scalability and does not even support the FRFP SFC aspects that this paper also targets. 

More practically, in~\cite{bolla2014dropv2}, the authors present an ETSI-support distributed VNF-supported infrastructure based on MANO framework~\cite{etsi2014network} to manage SDN-enabled switch (i.e., SDN node) energy consumption to meet regulatory and environmental standards. It targets the CPU energy consumption of the node and partially turns off some hardware components. Although this method is an available solution for practical scenarios, it should be carefully tuned to balance the trade-off between energy efficiency and function performance or one-demand SLAs. Moreover, unlike our solution, FRFP, it does not cover fog-supported SFC and fault tolerance failure recovery SDN-enabled switches. Furthermore, the authors in \cite{zhu2016joint} develop an energy-aware component SDN platform that targets data centers. This energy-aware component adopts priority-based Dijkstra for flow routing and exclusive scheduling across the network. This method has a big limitation, which backs to its traffic characteristics. In other words, it estimates neither the failure prevention nor the fault tolerance per SDN-enabled switch. To cover the limitation of the previous work, the authors in \cite{awad2017energy} present an energy-efficient routing solution in SDN by targeting the integral routing and discrete link rates as traffic characteristics. They solve the optimal MILP solution using two sub-optimal heuristics and validate it by single or multiple flows in a dynamic network. Overall, their contribution does not afford, by design and solution, joint failure recovery and fault tolerance of the fog-supported SDN-enabled switches and does not support SFCs variations in such a network that are addressed and validated in this paper.

Focusing on Fog computing appliances in SDNs/NFVs, there are limited works that target the routing in fog-supported SDNs/NFVs, such as \cite{truong2015software,luan2015fog, shojafar2016energy,tajiki2017joint}. In particular, the authors in \cite{truong2015software} address Fog computing over SDNs structures that preserve safety and non-safety services and are validated across two use cases: Data streaming and lane-change assistance services. The authors do not present any discussion about the SFC fault probability minimization, or failure recovery/prevention. In another work, in \cite{luan2015fog} the authors push the Fog Node to remain in edge to manage the on-demand location-based applications flows received from mobile users engaged in SDNs/NFVs and analyze the possible routing in such network. Unlike our method, it suffers from a lack of service chain management, fault-awareness, failure prevention and SDN-enabled switch energy minimization.

Moreover, recently in another work~\cite{shojafar2016energy}, the authors address the resource allocation and total energy minimization over the Fog Nodes by proposing a novel QoS-aware distributed and scalable scheduler. Although based on the authors’ claim that it can be applied in real-time services, it fails to address the chain of services when it faces fault and failure in such a dynamic network. Interestingly, our architecture, FRFP, can cover all the benefits of this method by covering all the limitations addressed. The most recent method similar to our current work is our previous work~\cite{tajiki2017joint} on SFC management in SDNs/NFVs. We present energy-aware resource reallocation SFC algorithms for SDNs. We allocate VNF to a set of flows and find several optimal and near-optimal solutions to optimize such network. Compared to our contribution, the paper~\cite{tajiki2017joint} has several limitations: (i) the presented routing algorithms do not exploit the capability of routing all flows simultaneously, i.e., it is impossible to reroute a flow considering the possible routes of other flows; and, (ii), we did not adopt the fog nodes to support fault probability minimization and failure recovery/prevention.

\section{The Proposed Architecture}\label{sec:refscenario}
In this section, we define the problem and assumptions in Sec.~\ref{subsec:probDefAsu} and provide a detailed discussion of the proposed architecture and its components in Sec.~\ref{sec:architecture}.

\subsection{Problem Definition and Assumptions}\label{subsec:probDefAsu}
    We consider multiple SDN-enabled routers/switches (referred as switches) with different fault probabilities that change during time slots. A central controller is connected to the switches to fetch the network information and configure the switches, using a southbound protocol to dynamically program the switches. There are several Fog Nodes in the network connected to the edge switches, and a maximum of one Fog Node is connected to each switch. We refer to the (Fog Node, switch) pair as a node throughout this paper. There are several types of servers in each Fog Node, and these have different rates of energy consumption. For a given Fog Node, the processing load cannot exceed a predefined threshold. Each Fog Node has a given processing capacity and can host several types of VNFs. the set of supported VNFs on each Fog Node is given. Each type of VNF requires a different processing capacity to process a unit of data, and the processing time for a VNF on different Fog Nodes for equal flow rates is the same. 
    
    Each flow needs to meet a set of VNFs along its path from the source to the destination switch (we refer to this requirement as the \textit{SFC requirement} of flows). The flow source, destination, set of required VNFs and rate are known. In addition, the end-to-end delay of transmission and the processing time of a flow should be less than a predefined threshold. We ignore queuing delay in the nodes. The traffic rate is dynamic and may change during the different time slots. 
    
    We define two different problems: i) recovery of the network in case of failure of one or more nodes, in such a way that the energy consumption of the network is minimized and the QoS requirement of the flows is satisfied; ii) periodic reconfiguration of the network in order to optimize the probability of a fault in selected paths for the active flows and reduce the overall energy consumption of the nodes.
    \begin{figure}[htpb]
	\begin{center}
		\includegraphics[width=0.9\textwidth]{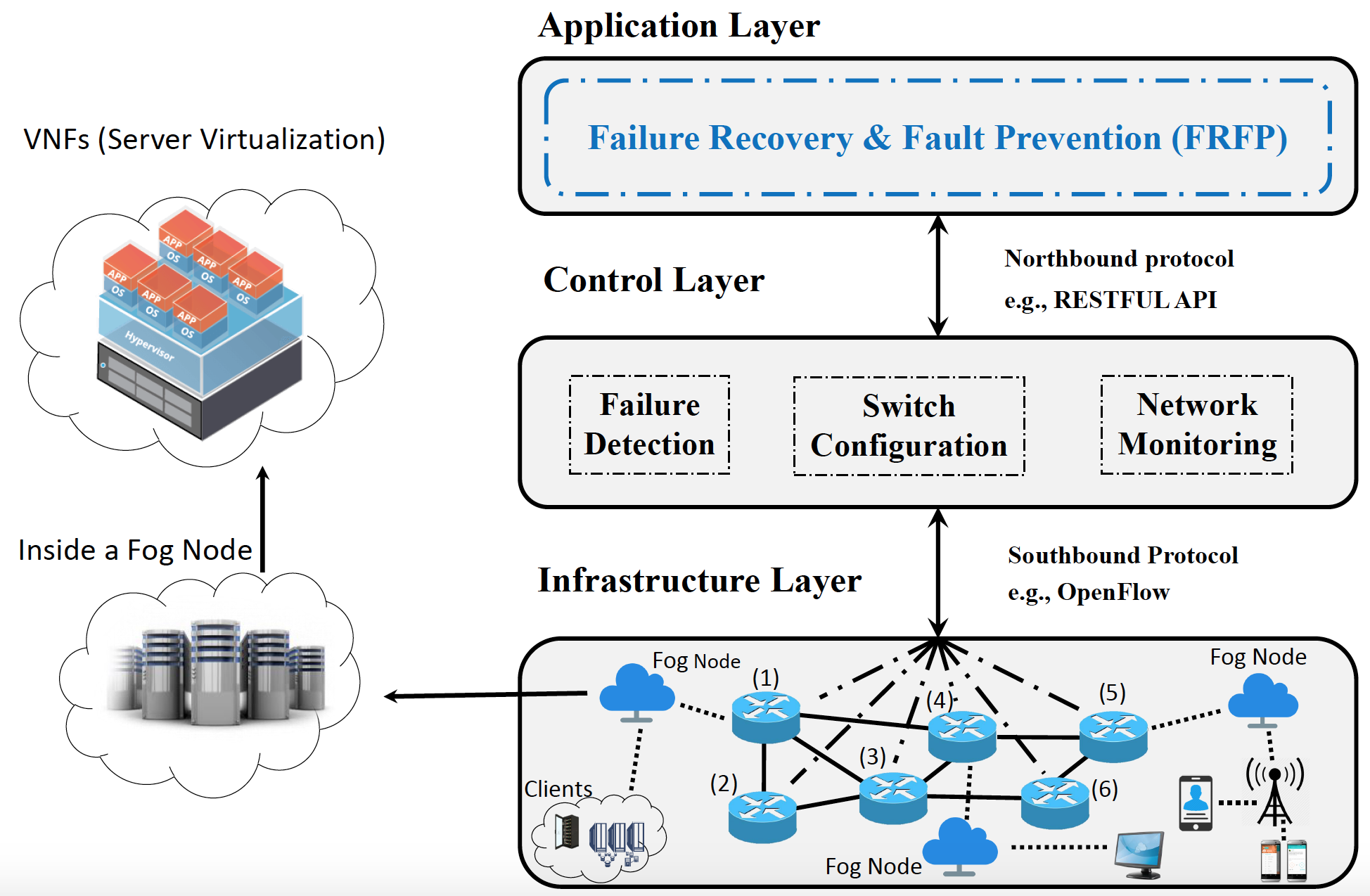}
		\caption{System Architecture: This architecture is based on the SDN principles and exploits server virtualization to run several VNFs on a single server. Each Fog Node contains one or more physical server.}
		\label{fig:Architecture}
	\end{center}
    \end{figure}
\subsection{Proposed Architecture}\label{sec:architecture}
        In this section, the proposed architecture and its components are presented. There are three layers in the proposed architecture: i) the Application Layer; ii) the Control Layer; and iii) the Infrastructure Layer. The Application Layer contains the resource assignment algorithms and routing protocols; in other words, the Application Layer is the brain in our architecture. The Control Layer consists of components that are required to: i) gather information about the network infrastructure or traffic patterns; and ii) force the switches and the Fog Nodes to operate based on the decisions that are made in the Application Layer. The proposed architecture is shown in Fig.~\ref{fig:Architecture} and has four components: Failure Recovery and Fault Prevention, Failure Detection, Switch Configuration, and Network Monitoring.

        \subsubsection{\textit{Failure Recovery and Fault Prevention (FRFP)}} This component periodically assigns network and Fog Node resources to flows, in order to reduce the energy consumption and simultaneously optimize the fault probabilities. In addition, if a switch undergoes a failure, this component re-assigns resources to the flows passing through the failed node in real time. To this end, FRFP considers three aspects: the QoS requirement of the flows, the energy consumption of the Fog Nodes, and the fault probability of the new resource assignments. This component belongs to the Application Layer. We formulate this component mathematically in Section~\ref{sec:formulation} and provide a corresponding fast heuristic algorithm in Section~\ref{sec:HFNR}. Three events activate FRFP: i) the arrival of a new flow; ii) failure in one or more nodes; and iii) the end of the timer interval (periodically). The input of this component in case of the arrival of a new flow is the specification of the flow (e.g., source, destination, rate, and set of required VNFs) and the current state of the nodes (e.g., utilization of links and Fog Nodes). In case of a failure, the input is the specification of the flows passing through the failed node(s) and the current state of the network. It should be mentioned that if a node (switch or Fog Node) fails, the network topology and/or the list of supported VNFs (in each Fog Node) may change. For periodic re-assignment of resources, the specification of all flows along with the current state of the network is the input of the FRFP.
        
        \subsubsection{\textit{Failure Detection}}
        This component is responsible for the detection of node failures or crashes. In case a switch or Fog Node fails, this component sends the new network topology and the set of supported VNFs for each Fog Node to the FRFP component. This component is part of the Control Layer. 
        
        \subsubsection{\textit{Switch Configuration}} The task of this component is to apply the decisions made in the Application Layer to the switches, via the configuration of the switch forwarding tables. In other words, this component allocates/reallocates resources to flows by rescheduling the forwarding tables. This component is part of the Control Layer.
        
        \subsubsection{\textit{Network Monitoring}} This component continuously monitors the network traffic by gathering information from switches and sending the information to the Application Layer. In other words, this component can obtain the current flow matrix and network topology by querying the SDN switches. This component is part of the Control Layer.
   
\section{System Model}\label{sec:model}
    In this section, we describe the notations used in this paper. Table~\ref{tab:notation} defines the symbols, presents their type and units, their appearances in the equations and provides a brief description of them.
    \begin{table*}[t]
    	\caption{Main Notation.}\label{tab:notation}
        \centering
        \scriptsize
        \rowcolors{2}{gray!25}{white}
        \begin{tabular}{|c|l|l|l|}	
    	\hline
    	    \textbf{Symbol} & \textbf{Definition} & \textbf{Type - Unit} & \textbf{Appears in Eq.} \\\hline\hline
    	    \multicolumn{4}{|c|}{\textbf{Input Parameters}}\\\hline
    		$\mathcal{N}$ & Set of switches, $\abs{\mathcal{N}}\triangleq N$ &- &-\\\hline
    		$\mathcal{F}$ & Set of flows, $\abs{\mathcal{F}}\triangleq F$ &- &-\\\hline
    		$\mathcal{X}$ & Set of functions, $\abs{\mathcal{X}}\triangleq X$ &- &-\\\hline
    		$E$ & Number of links & Integer - [units] &-\\\hline
    		$\mathcal{T}$ & Total number of time slots & Integer - [units] &-\\\hline
    	    $\Psi$ & Maximum number of required functions for each flow & Integer - [units] &-\\\hline
    		$B_{(i,j)}$ & Matrix of link bandwidth between $i$-th and $j$-th switches & Continues - [Mb/s]&\eqref{eq7}\\ \hline
    		$D_{(i,j)}$ & Links propagation delay & Continues - [ms]&\eqref{eq12}\\\hline
    		$\mu$ & Maximum link/Fog-Node utilization & Continues - [units]&\eqref{eq7}\\\hline
    		$MT$ & Maximum tolerable joint failure probability & Continues - [ms]&\eqref{eqFault5},\eqref{eqFault1}\\\hline
    		$T^f$ & Maximum tolerable delay of flow & Continues - [ms]&\eqref{eqPropageDelayDefinition}\\\hline
    		$TP_x$ & Processing time of VNF $x$ for one unit of data  & Continues - [ms]&\eqref{eqPropageDelayDefinition}\\\hline
    		$C^f(t)$ &  Bandwidth requirement matrix for the $f$-th flow in time slot $t$ & Continues - [Mb/s]&\eqref{eq7}\\ \hline
    		$s^{f}$ & Vector of source switch for the $f$-th flow & Integer - [units]&\eqref{eq8} \\ \hline
    		$d^{f}$ & Vector of destination switch for the $f$-th flow & Integer - [units]&\eqref{eq8},\eqref{eqFault1},\eqref{eqFault2}\\ \hline
    		${FP}_{x}$ & Required processing for the $x$-th function & Continuous - [units]&\eqref{eq6} \\\hline
    		${NC}_{i}$ & Nodes processing capacity for the $i$-th node & Continuous - [units]&\eqref{eq6}\\\hline
    		$FN_{(i,x)}$ & Function $x$ associated with $i$-th node & Binary - [units]&\\ \hline
    		$R^f_{x}(t)$ & Requested functions for the $f$-th flow in time slot $t$& Binary - [units]& \eqref{eq4}\\ \hline
    		$\mathcal{E}_{i}$ & Power consumption for $i$-th node& Continuous - [W]& \eqref{eqEnergy} \\\hline
    		$p_{i}(t)$ & Failure probability for switch $i$ in time slot $t$& Continuous - [units]&\eqref{eqFault1}\\\hline\hline
    		\multicolumn{4}{|c|}{\textbf{Variables}}\\\hline
    		$\mathcal{P}_{r}(t)$ & Fault probability for path ID $r$ in time slot $t$& Continuous - [units] &\eqref{eqFault5}\\\hline
    		$A_{(i,j)}^f(t)$ & Network resource assignment matrix between $i$-th and $j$-th switches with the flow $f$ in time slot $t$& Binary - [units]&\eqref{eqNetSidEfct},\eqref{eqFault1},\eqref{eqFault2},\eqref{eq3},\eqref{eq7},\eqref{eq8},\eqref{eq9},\eqref{eq12}\\\hline
    		$U_{(i,x)}^f(t)$ & Used services for the $i$-th switch with the flow $f$ that runs the function $x$ in time slot $t$& Binary - [units]&\eqref{eq2},\eqref{eq3},\eqref{eq4},\eqref{eq5},\eqref{eq6},\eqref{eq10}\\\hline
    		$T_d^f$ & Maximum tolerable propagation delay of flow $f$ &Continues - [ms]&\eqref{eqPropageDelayDefinition},\eqref{eq12}\\\hline
    		$O_{i}(t)$ & ON/SLEEP nodes in time slot $t$& Binary - [units]&\eqref{eqEnergy},\eqref{eq10}\\\hline
    		$\mathcal{Z}^f(t)$ & Index of selected path for flow $f$ in time slot $t$& Integer - [units] &\eqref{eqFault2},\eqref{eqFault4}\\\hline
    		$E(t)$ & Total energy Consumption in time slot $t$& Continuous - [J]&\eqref{eqEnergy},\eqref{eqObjective}\\\hline
    		$NS(t)$ & Network side-effect of flow rerouting in time slot $t$& Integer - [units] &\eqref{eqObjective}\\\hline
    		$J_i^f(t)$ &Path allocation vector $i$ used for flow $f$ in time slot $t$&Binary - [units]&\eqref{eqFault3}, \eqref{eqFault4}, \eqref{eqFault5}\\\hline
    	\end{tabular}
    \end{table*}
    Let $N$ be the number of SDN-enabled switches. We represent the network topology with a matrix $B_{N\times N}$ where $B_{(i,j)}$ denotes the capacity of the link from the switch $i$ to the switch $j$. Similarly, the propagation delay of links is modeled via matrix $D_{N\times N}$ where $D_{(i,j)}$ denotes the propagation delay of the link from the switch $i$ to the switch $j$. Let $F$ be the number of flows in the network. In order to simplify the understanding of the notations, a sample for the each element of the proposed notations is presented. For example, for the topology illustrated in Fig. \ref{fig:Architecture}, the matrix $B$ and $D$ are as follows ($N=6$):
     \begin{table*}[!h]
        \centering
        \begin{tabular}{c c}
           $ B=
              \begin{bmatrix}
                0 & \textcolor{blue}{B_{(1,2)}} & \textcolor{blue}{B_{(1,3)}} & \textcolor{blue}{B_{(1,4)}} & 0 & 0\\
                \textcolor{blue}{B_{(2,1)}} & 0 & \textcolor{blue}{B_{(2,3)}} & 0 & 0 & 0\\
                \textcolor{blue}{B_{(3,1)}} & \textcolor{blue}{B_{(3,2)}} & 0 & \textcolor{blue}{B_{(3,4)}} & 0 & \textcolor{blue}{B_{(3,6)}}\\
                \textcolor{blue}{B_{(4,1)}} & 0 & \textcolor{blue}{B_{(4,3)}} & 0 & \textcolor{blue}{B_{(4,5)}} & 0\\
                0 & 0 & 0 & \textcolor{blue}{B_{(5,4)}} & 0 & \textcolor{blue}{B_{(5,6)}}\\
                0 & 0 & \textcolor{blue}{B_{(6,3)}} & 0 & \textcolor{blue}{B_{(6,5)}} & 0
              \end{bmatrix}$
            &$ D=
              \begin{bmatrix}
                \infty & \textcolor{blue}{D_{(1,2)}} & \textcolor{blue}{D_{(1,3)}} & \textcolor{blue}{D_{(1,4)}} & \infty & \infty\\
                \textcolor{blue}{D_{(2,1)}} & \infty & \textcolor{blue}{D_{(2,3)}} & \infty & \infty & \infty\\
                \textcolor{blue}{D_{(3,1)}} & \textcolor{blue}{D_{(3,2)}} & \infty & \textcolor{blue}{D_{(3,4)}} & \infty & \textcolor{blue}{D_{(3,6)}}\\
                \textcolor{blue}{D_{(4,1)}} & \infty & \textcolor{blue}{D_{(4,3)}} & \infty & \textcolor{blue}{D_{(4,5)}} & \infty\\
                \infty & \infty & \infty & \textcolor{blue}{D_{(5,4)}} & \infty & \textcolor{blue}{D_{(5,6)}}\\
                \infty & \infty & \textcolor{blue}{D_{(6,3)}} & \infty & \textcolor{blue}{D_{(6,5)}} & \infty 
              \end{bmatrix}$
        \end{tabular}
    \end{table*}
            
    The source and destination of flows are determined by vectors $s^f$ and $d^f$, respectively. The matrix $A_{N\times N\times F}(t)$ is the assignment of network resources (links) to the flows, such that if $A_{(i,j)}^f(t_1)= 1$, then the flow $f$ passes the link $i\rightarrow j$ in time slot $t_1$. If we set $f=1$, $t=t_1$, $s^1=1$, and $d^1=2$, then the matrix $A^1(t_1)$ becomes as follows:
    \begin{center}
                   $ A^1(t_1)=
              \begin{bmatrix}
                0 & 0 & \textcolor{blue}{1} & 0 & 0\\
                0 & 0 & 0 & 0 & 0\\
                0 & 0 & 0 & 0 & \textcolor{blue}{1}\\
                0 & \textcolor{blue}{1} & 0 & 0 & 0\\
                0 & 0 & 0 & \textcolor{blue}{1} & 0
              \end{bmatrix},$
    \end{center}
    where $s^1=1$ indicates that the source of the flow is the switch number 1. Therefore, we should trace the path from the first row of $A^1(t_1)$. As can be seen, the third element of $A^1(t_1)$ in the first row is one which means that the flow will leave the switch 1 toward the switch 3. At this point, the third row of $A^1(t_1)$ should be checked. Since the $5^{th}$ column of the third row is 1, the flow will leave switch number 3 to reach the switch number 5. The flow will go to switch number 4 because the fourth element of row 5 in matrix $A^1(t_1)$ is one. Finally, since the second column of the forth row is one, the flow will go to switch number 2. Note that we consider loop-free routing, i.e., nodes and links cannot be used twice in the routing of a flow. We will enforce this behavior with specific constraints in our formulation.
    
    Considering $X$ different VNFs, each flow can request at most $\Psi\leq{X}$ VNFs. The set of requested VNFs for each flow is shown by matrix $R_{F\times X}$. Therefore, if $R_{x}^f(t)$ is 1, then the VNF $x$ is requested for the flow $f$ in time slot $t$. As an example, considering $X=4$ and $\Psi=3$ in time slot $t$ for flow $f=1$, the matrix $R^1_4(t)$ is as follows:
    \[R^1_4(t)=
      \begin{bmatrix}
        {\textcolor{blue}{1}} & 0 & 0 & {\textcolor{blue}{1}}
      \end{bmatrix},\ \ \ 
    \]
            
   The first and the last elements of $R^1_4(t)$ are one, meaning that the flow should deliver service from VNF number one and three.
   The matrix $C^{f}(t)$ specifies the flow rates in time slot $t$. The $i^{th}$ row of this matrix defines the traffic rate requirement of the $i^{th}$ flow. Vector $T^f$ specifies the maximum processing and communication delay that the flow can tolerate. The vector $TP_{1\times X}$ specifies the processing time of one unit of data over each VNF; e.g., $TP_x=3$ [ms] means that VNF $x$ needs 3 [ms] to process one [unit] of data. $T^F_d$ is the maximum tolerable propagation delay\footnote{Note that in this paper we do not consider the queuing delay} of the flows.     
    
    The required processing capacity of each VNF for a unit of flow rate is expressed by the vector $FP_X$, where $FP_x$ specifies the required processing capacity of VNF $x\in \mathcal{X}$. Therefore, the VNF $x$ will require a processing capacity $FP_x\cdot C^{f}(t)$ to process the flow $f$ with rate $C^{f}(t)$ in time slot $t$. The vector $NC_{1\times N}$ identifies the processing capacities for each Fog Node. The VNFs associated with each Fog Node are identified by matrix $FN_{N\times X}$. Therefore, $FN_{(i,x)}$ specifies whether VNF $x$ is supported by Fog Node $i$ or not. We consider a Fog Node connected to each switch. If no Fog Node is connected to switch $i$, then $\sum_{x=1}^X FN_{(i,x)}=0$. $U^F_{N\times X}(t)$ denotes the assignment of the VNFs and Fog Nodes to the flows in time slot $t$. If $U^f_{(i,x)}(t)$ is 1, then flow $f$ receives service from VNF $x$ on Fog Node $i$ in time slot $t$. Taking $f=1$, an example matrix $U^1(t_1)$ is:
        \[U^1(t_1)=
          \begin{bmatrix}
            0 & 0 & {\textcolor{blue}{1}} & 0\\
            0 & 0 & 0 & 0\\
            0 & 0 & 0 & 0\\
            0 & {\textcolor{blue}{1}} & 0 & 0\\
            0 & 0 & 0 & 0\\
            0 & 0 & 0 & 0
          \end{bmatrix}.\]
          
\begin{figure*}[!htbp]
    \centering
        \begin{subfigure}{\textwidth}
    	    \centering
    		\includegraphics[width=1\linewidth]{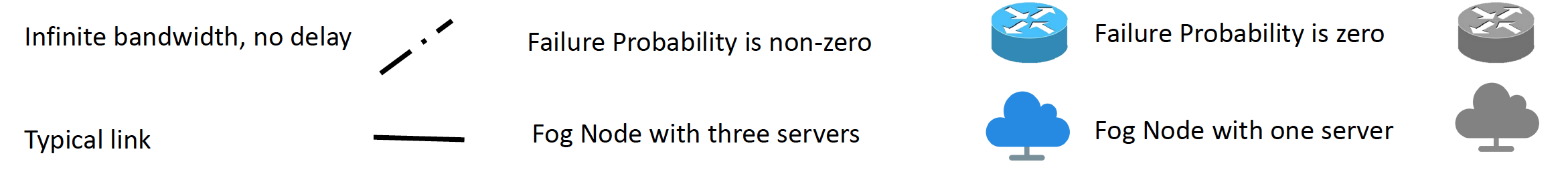}
    	\end{subfigure}
    	\begin{subfigure}{0.45\textwidth}
    	    \centering
    		\includegraphics[width=1\linewidth]{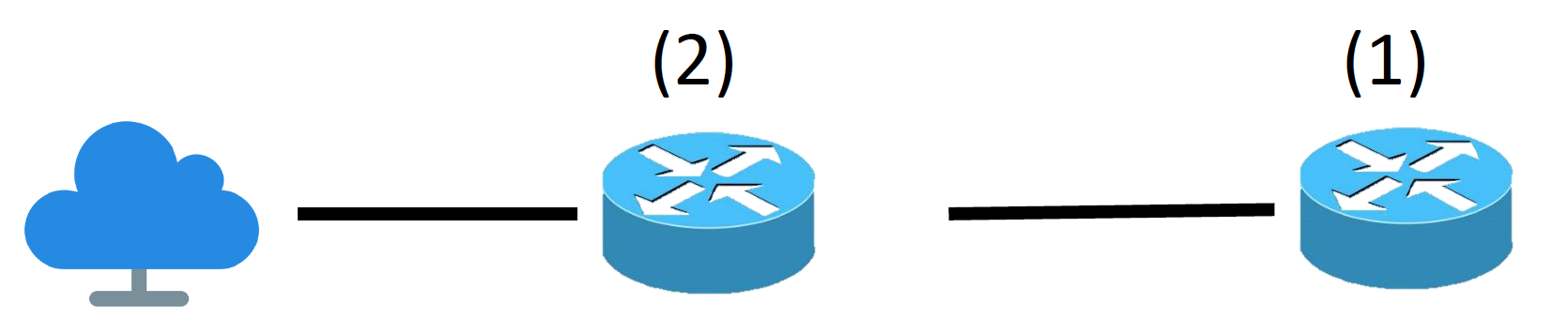}
        	\caption{Before substitution.}
        	\label{fig:multiServToFog1}
    	\end{subfigure}
    	\begin{subfigure}{0.45\textwidth}
    	    \centering
    		\includegraphics[width=1\linewidth]{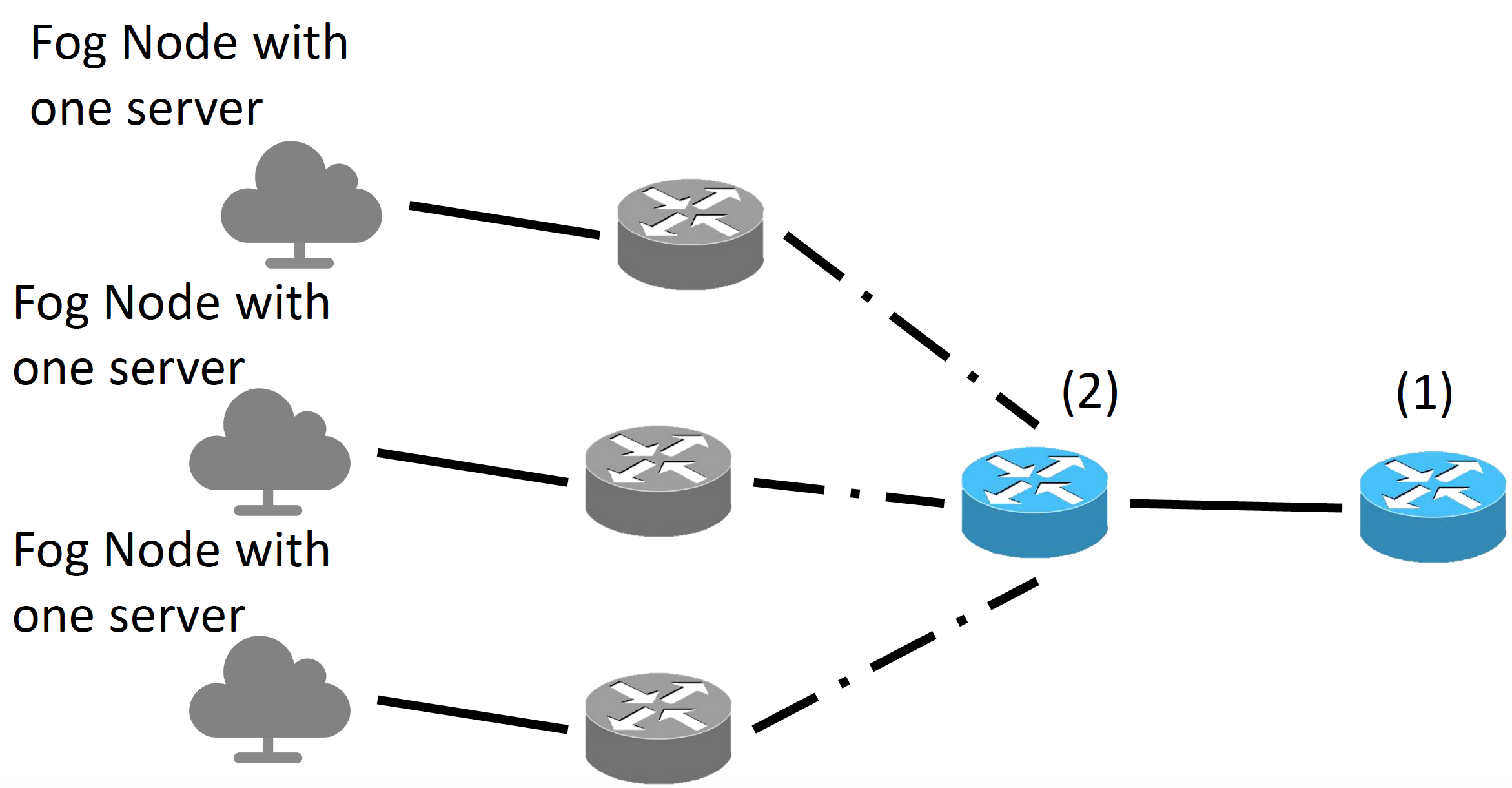}
        	\caption{After substitution.}
        	\label{fig:multiServToFog2}
    	\end{subfigure}
    	\caption{Substituting a Fog Node with three server with three Fog Nodes each one containing one server.}
        \label{fig:multiServToFog}
    \end{figure*}

    In the first row of $U^1(t_1)$, the third element is one, implying that, the flow uses the VNF number three which is hosted by the Fog Node number 1. Similarly, since the second element of the fourth row is one, the VNF 2 will be delivered to the flow in Fog Node~4.
    In our formulation, we substitute a Fog Node that has more than one server with several nodes where each node has a Fog Node which has only one server. Figure~\ref{fig:multiServToFog} illustrates an example of how a Fog Node which has three servers can be substituted with three Fog Nodes where each Fog Node has only one server.
    
    Each Fog Node has two different modes: ON and IDLE. If no VNF is active on a Fog Node, the Fog Node goes to IDLE mode, otherwise, the Fog Node is in ON mode. The energy consumption of Fog Nodes in ON mode are stated using vector $\mathcal{E}_{1\times N}$, where $\mathcal{E}_i, i \in \mathcal{N}$ specifies the energy consumption of Fog Nodes $i$. In the IDLE mode, the energy consumption is a fraction of full-rate energy consumption $\mathcal{E}-(\delta\cdot \mathcal{E})$. The current state of Fog Nodes is specified by $O_{1 \times N}(t)\in \{0,1\}$ where $O_i(t_1)=1$ means that the Fog Node $i$ is in ON mode in time slot $t_1$. The variable $\mu$ states the maximum link and Fog Node utilization. Focusing on the fault prevention, the end-to-end fault probability of routing in the network for flow $f$ should be less than a predefined threshold $MT$. The fault probability of switches is stated using $p_N(t)$, e.g., $p_i(t_1)$ specifies the fault probability of switch $i$ in time slot $t_1$. 
    
\section{Problem Formulation}\label{sec:formulation}
    In this section, we present the SFC-aware congestion control system with the goal of minimizing energy consumption of the Fog Nodes. Also, we focus on the fault probability of the selected path for each flow. Hence, it guarantees the required service functions of each flow to be delivered via the selected path. Additionally, it ensures the end-to-end fault probability of the selected path for each flow to be less than a predefined threshold. In this section, we initially detail each set of constraints, and then we provide the complete formulation. 
    
\subsection{QoS, Routing, and Delay}\label{sec:QoSRoutingDelay}
    Focusing on the link capacity, constraint~\eqref{eq7} checks the link capacity between each pair of switches.
    \begin{align}
            &\sum^F_{f=1}{\left(A_{(i,j)}^f(t) \ \cdot \ C^f(t)\right)}\leq \mu \cdot  B_{(i,j)},\forall i,j\in \mathcal{N}.\label{eq7}
    \end{align}
    
    Focusing on the flow conservation, constraint~\eqref{eq8} presents the flow management limitations. The first equality states that a flow leaves its source only once. The second equality imposes states that a flow enters a destination and does not leave it. Also, the third equality forces the input and output of each node to be equal (i.e., except for the source and destination).
    \begin{align}
    &\sum^N_{j=1}{A_{(i,j)}^f(t)}-\sum^N_{j=1}{A^f_{(j,i)}(t)}=
        \begin{cases}
            1 & \text{if } i=s^f\\
            -1 & \text{if } i=d^f\\
            0 & \text{Otherwise}\\
        \end{cases},
        \forall f\in \mathcal{F},\ \forall i\in\mathcal{N},\label{eq8} 
    \end{align}
    
    Constraint~\eqref{eqPropageDelayDefinition} denotes the maximum communication delay which is the propagation delay of the path from the source switch to first selected Fog Node, from the first selected Fog Node to the other Fog Nodes (if applicable), and from the last selected Fog Node to the destination switch.
    \begin{align}
        &T^f_d=T^f-\sum^X_{x=1}{\left(TP_x\cdot R^f_x(t)\cdot C^f(t)\right)},\  \forall f\in \mathcal{F}\label{eqPropageDelayDefinition}
    \end{align}
    
    Focusing on the propagation delay, constraint~\eqref{eq12} is used to control the propagation delay for each flow. In order to prevent loops for each flow, constraint~\eqref{eq9} is applied. 
        \begin{align}
            &\sum^N_{i=1}{\sum^N_{j=1}{\left(A^f_{(i,j)}(t-1)\ \cdot\ D_{(i,j)}\right)}}\leq T^f_d,\ \forall f\in \mathcal{F},\label{eq12} \\
            &\sum^N_{j=1}{A_{(i,j)}^f(t)}\leq 1,\ \forall i\in \mathcal{N},\ \forall f\in \mathcal{F},\label{eq9} \\
            &A_{(i,j)}^f(t)\in \left\{0,\ 1\right\},\ \forall i,j\in \mathcal{N},\ \forall f \in \mathcal{F},\ \forall t \in \mathcal{T}.\nonumber
        \end{align}
    
\subsection{Faults}\label{sec:faults}
       In order to minimize the effect of faults in the network, we formulate the probability of a fault in each network path. To this end, we calculate the survival probability of the path and then calculate its fault probability. Let $p_{i}(t)$ denote the fault probability for switch $i$ in time slot $t$ under independent failure assumptions. Therefore, the switch $i$ will survive with probability $1-p_{i}(t)$. Thus, the survival probability of path $r$ is $\prod_{\forall i\in r}{(1-p_{i}(t))}$ in time slot $t$ and consequently the fault probability of path $r$ in time slot $t$ is $1-\prod_{\forall i\in r}{(1-p_{i}(t))}$.
        \begin{align}
            a\triangleq\left(1-p_{d^f}(t)\right)\label{eq:aDef}
        \end{align}
        In this way, $a$, which is defined in constraint~\eqref{eq:aDef}, denotes the survival probability of the destination switch $d^f$ of flow $f$ in time slot $t$.
        \begin{align}
            &b\triangleq\prod_{\forall i \in \mathcal{N}}{\left((1-p_i(t))\cdot \sum_{j=1}^N{A_{(i,j)}^f(t)}\right)}\label{eq:bDef}
        \end{align}
        Correspondingly, $b$ (in constraint~\eqref{eq:bDef}) denotes the survival probability in the rest switches of the selected path $r$ for flow $f$ in time slot $t$. Therefore, the survival probability of the selected path $r$ for flow $f$ in time slot $t$ will be $(a\times b)$. Consequently, the fault probability of the selected path $r$ in time slot $t$ is $1-(a\times b)$.
        
        \noindent We define constraint \eqref{eqFault1} to guaranty the probability of end-to-end switching fault be less than a predefined maximum tolerable. 
        \begin{align}\small
            & 1-\left(\prod_{\forall i \in \mathcal{N}}{\left((1-p_i(t))\cdot \sum_{j=1}^N{A_{(i,j)}^f(t)}\right)}\cdot\left(1-p_{d^f}(t)\right)\right)
            ~\leq MT,\  \forall f\in \mathcal{F}, \forall t\in \mathcal{T} \label{eqFault1}
        \end{align}
        However, constraint~\eqref{eqFault1} is a non-linear constraint and next we describe hot to handle it. 
        Without loss of generality, we substitute the constraint~\eqref{eqFault1} with Eqs. \eqref{eqFault2}-\eqref{eqFault5} to keep the formulation in linear form. To this end, we generate an ID to differentiate between different paths. The paths with the same fault probabilities have the same ID. Focusing on the path selection, $Z^f(t)$ identifies the ID that is assigned to the path selected for flow $f$ in time slot $t$. If the path which is selected for flow $f$ includes switch number $i$, then $2^{(i-1)}$ is added to $\mathcal{Z}^f$. Consequently, the paths which contain the same set of switches, are considered to have the same ID. This happens because $\mathcal{Z}^f(t)$ is used to differentiate between paths with different fault probability and those paths that have the same set of switches have similar fault probability. E.g., if flow $f_1$ passes throw switch $i$ in time slot $t$ but flow $f_2$ does not pass throw that switch in time slot $t$ then $Z^{f_1}(t)\neq Z^{f_2}(t)$. 
        
        \noindent Let $ID^f_{i}(t)\triangleq\sum_{j=1}^N{A^f_{(i,j)}(t)}\cdot 2^{(i-1)}$ denote the ID number sets for SDN-enabled switch $i$ for flow $f$ in time slot $t$. In another words, if flow $f$ passes throw switch $i$ in time slot $t$, then the $ID^f_{i}(t)$ will be $2^{(i-1)}$, otherwise it will be zero. Therefore, $Z^f(t)$ can be calculated as follows:
       \begin{align}
            &\mathcal{Z}^f(t)\triangleq\sum_{i=1}^N{\left(\sum_{j=1}^N{A^f_{(i,j)}(t)}\cdot 2^{(i-1)}\right)}+2^{(d^f-1)},
                ~\forall f\in \mathcal{F}, \forall t\in \mathcal{T}\label{eqFault2}
        \end{align}
        According to Section~\ref{sec:model}, if the destination of flow $f$ is $d^f$, then $A^f_{(d^f,j)}(t)$ will be 0. Hence, $ID^f_{d_f}(t)$ is always zero for destination switch. Because of this, $2^{(d^f-1)}$ is added to constraint~\eqref{eqFault2} to include the impact of the fault probability of the destination switch on the selected path $r$ for flow $f$. In brief, for a network with $N$ switches, the value of $\mathcal{Z}^f(t)$ can be a number between 0 - $2^N$.
        \begin{align}
            &\sum_{r=1}^{2^N}{J_r^f(t)}=1,\ \forall f\in \mathcal{F}\label{eqFault3}
        \end{align}
        Variable $J^f_r(t)$ specifies whether $\mathcal{Z}^f(t)$ is $r$ or not, e.g., $J^f_r(t)=1$ means $\mathcal{Z}^f(t)=r$. Note that there are several different paths that have equal ID value (i.e., all paths that have same set of switches (with different ordering) have identical path ID). Equation \eqref{eqFault3} guarantees that in each time slot, only one ID is assigned to flow $f$.
        \begin{align}
            &\sum_{r=1}^{2^N}{(J_r^f(t)\times r)}=\mathcal{Z}^f,\ \forall f\in \mathcal{F}\label{eqFault4}
        \end{align}
        Since $Z^f(t)$ is the ID of the selected path which is captured from matrix $A(t)$, constraint~\eqref{eqFault4} checks the consistency of the formulation. Finally, constraint~\eqref{eqFault5} states the condition under which end-to-end fault probability of the selected path is \textit{lower} than a predefined threshold.
        \begin{align}
            &\sum_{r=1}^{2^N}{(J_r^f(t)\times \mathcal{P}_r(t))}\leq MT,\ \forall f\in \mathcal{F}
            \label{eqFault5}
        \end{align}
        
\subsection{Service Function Chaining (SFC)}\label{sec:SFC}
    Regarding the constraints in ~\eqref{eq2}-~\eqref{eq6}, some remarks are in order. The constraint~\eqref{eq2} indicates that each flow crosses a valid function chain while passing through the switches.
    \begin{align}
        &\sum^N_{i=1}{U_{(i,x)}^f(t)}\geq R^f_{x}(t),\ \forall x\in \mathcal{X},\ \forall f\in \mathcal{F},\label{eq2} 
    \end{align}
    
    \noindent Moreover, constraint~\eqref{eq3} imposes the service delivery only on crossed nodes. Constraint~\eqref{eq4} checks whether the requesting function is supported on the specified node.
    \begin{align}
        &\sum^N_{i=1}{A_{(i,j)}^f(t)}\geq U^f_{(j,x)}(t),\ \forall x\in \mathcal{X},\ \forall j\in \mathcal{N}-\{s^f\},\ \forall f\in \mathcal{F},\label{eq3} \\
        &U_{(i,x)}^f(t)\leq FN_{(i,x)},\ \forall f\in \mathcal{F},\ \forall i\in \mathcal{N},\ \forall x\in \mathcal{X},\label{eq4}
    \end{align}
    
    \noindent Constraint~\eqref{eq5} prevents using a service function more than once for each flow. Constraint~\eqref{eq6} controls the capacity of nodes providing a service.
    \begin{align}
        &\sum^N_{i=1}{U_{(i,x)}^f(t)}= 1,\ \forall f\in \mathcal{F},\:\forall x\in \mathcal{X},\label{eq5} \\
        &\sum^F_{f=1}{\sum^X_{x=1}{\left(U_{(i,x)}^f(t)\ \cdot \  FP_x\ \cdot \ C^f(t)\right)}}\leq {NC}_{i},\ \forall i\in \mathcal{N},\label{eq6}\\
        &U^f_{(i,x)}(t) \in \left\{0,\ 1\right\},\ \forall i\in \mathcal{N},\ \forall f \in \mathcal{F},\ \forall x \in \mathcal{X},\ \forall t \in \mathcal{T}.\nonumber
    \end{align}
        
    \subsection{Energy and Network Side-effect}\label{sec:EnergySide-effect}
        Let $E(t)$ be the energy consumption of the network in time slot $t$ which is calculated as
        \begin{align}
            &E(t)=\sum^N_{i=1}O_i(t)\:\cdot \:\mathcal{E}_i ,\label{eqEnergy} 
        \end{align}
        and the network side-effect (i.e., the number of forwarding table elements that should be changed to apply the new configuration) of flow rerouting in time slot $t$ as
        \begin{align}
          &{NS}(t)=\sum^N_{i=1}{\sum^N_{j=1}{\sum^F_{f=1}{\abs{A^f_{(i,j)}(t)-A^f_{(i,j)}(t-1)}}}} ,\label{eqNetSidEfct} 
        \end{align}
      
        \begin{align}
            &\left(1+F\cdot  X\right){O}_i(t)\geq \sum^F_{f=1}{\sum^X_{X=1}{U_{(i,x)}^f(t)}},\ \forall i\in \mathcal{N}, \label{eq10}
        \end{align}
       Constraint~\eqref{eq10} specifies which Fog Nodes must be ON (those that deliver at least one service to the flow).  
       
    \subsection{Overall Formulation}\label{sec:overall}
    The OPTIMAL FOG-SUPPORTED ENERGY-AWARE SFC (OFES) rerouting problem in FRFP architecture presented in Fig.~\ref{fig:Architecture}. which aims at minimizing jointly the energy and side-effect of the SDN-enabled switches at each time slot $t$. In another words, the objective function \eqref{eqObjective} is to optimize both the number of Fog Nodes that are required to be turned ON and the network side-effect of flow rerouting in time slot $t$. OFES is formulated as follows:
    \begin{align}
        &\min_{O,A}\left[\alpha\cdot E(t)+\beta\cdot{NS}(t)\right] ,\label{eqObjective}\\\nonumber
        &\text{subject to:}\\\nonumber
        &\text{QoS, Routing, and Delay} & \eqref{eq7}-\eqref{eq9}\\\nonumber
        &\text{Fault} & \eqref{eqFault2}-\eqref{eqFault5}\\\nonumber
        &\text{SFC} & \eqref{eq2}-\eqref{eq6}\\\nonumber
        &\text{Energy and Network Side-effect} & \eqref{eqEnergy}-\eqref{eq10}
    \end{align}
under control variables: $A_{(i,j)}(t) \in \{0,1\}$, $O_i(t) \in \{0,1\}$ and $U_{(i,j)}^f(t) \in \{0,1\}$. Moreover, $\alpha$ and $\beta$ $\in$ [0,1], where $\alpha+\beta=1$ are assigned weight factors that tune the desired trade-off between SDN consumed energy and SDN side-effect facing with failures and faults. 
   
\section{Heuristic Fog-supported Energy-aware SFC rerouting algorithm (HFES)}\label{sec:HFNR}
    Since the optimal solution is very challenging and complex to be solved even for instances of small-size real-time network reconfigurations, we propose the Heuristic fog-supported Energy-aware SFC rerouting algorithm called (HFES) to practically tackle it. \textbf{Algorithm \ref{alg:HFES}} reports the HFES pseudocode. The HFES algorithm is a recursive algorithm which guaranties that the probability of fault in the selected path is less than a predefined threshold.
    
    	In detail, in line 1 of the \textbf{Algorithm \ref{alg:HFES}}, we check the fault probability $(1-PR)$ of the selected path to be less than a predefined threshold $MT$. Note that we consider $PR$ as survival probability of a selected path that can be between 0 to 1. When  $PR==-1$, it means that the current path should not be traversed due to exceeding the predefined threshold $MT$ of the fault rate (i.e., it stands as stopping criteria in the HFES recursive algorithm). The solution is found if all of the required VNFs are met: $R==\emptyset$, and the final state of the selected path is the destination of the flow: $(CN==d)$. On the other hand, if all required VNFs are met, but the last switch of the selected path is not the destination of the flow (line 3) (i.e., the flow is unable to reach to the desired destination), the algorithm finds the next hop $NH$ of the shortest path from the current state to the flow destination and adds it to the selected path by invoking algorithm HFES with new inputs (lines 4--12). In detail, the algorithm removes nodes that are met to prevent loop using Remove function (line 6). In line 7 we calculate the survival probability of considering $NH$ as the last hop of the selected path $CP$. The algorithm uses the HFES recursively in order to check whether a valid solution is found or not (line 8).
    Lines 13-22 handle the cases where all of the required VNFs are not delivered to the flow $(R\neq \emptyset)$. In this case, a node which is directly connected to the current node $CN$ and has the minimum energy consumption is selected as the next hop $NH$ (line  15). Thereafter, the flow receives services from VNFs that are active in the current node $CN$ (line 16). These steps are iterated in HFES algorithm until $NH$ be empty (lines 17-23). The goal of HFES is to provide a path with a guaranteed fault probability while minimizing the length of the paths and minimizing Fog Nodes energy consumption (line 25). 
	\begin{algorithm}[!htbp]
    	\caption{Pseudo-Code of the HFES algorithm}
    	\label{alg:HFES}
    	\small
    	\allowdisplaybreaks
    	\begin{algorithmic}[1]
        	\INPUT{CN, d, R, B, NC, FP, p, PR,CP}        	\Statex{\texttt{CP: Chosen Path}}
        	\Statex{\texttt{CN: Current Node}} 
        	\Statex{\texttt{PR: Survival probability of path}}
        	\OUTPUT{$<$CP,\:PR$>$}
        	\If{PR$\geq$ 1-MT}
        	    \Return{$<$CP,\:-1$>$}
        	\ElsIf{R==$\emptyset$\ and CN==d}
        	    \Return{$<$CP,\:0$>$}
        	\ElsIf{R==$\emptyset$}
        	    \Do
                 \State{NH=NextHubShortestPath(CN,d,B);}
        	        \State{B$^\prime$=Remove(B,NH);}
        	        \State{PR$^\prime$=PR$\times$(1-p$_{NH}$);}
        	        \State{[CP$^\prime$, PR]=HFES(NH,d,R,B$^\prime$,NC,FP,p,PR$^\prime$,CP$^\prime$);}
        	        \If{PR$\neq$ -1}
        	            \State{CP=CP$^\prime$;}
        	        \EndIf
        	    \doWhile{NH$\neq\emptyset$}
        	\Else
        	    \Do
        	        \State{NH=EnergyAwareNextHub(CN,d,B,R);}
                     \State{[R$^\prime$, NC$^\prime$]=DeliverService(R,NH,NC,FP);}
                     \State{B$^\prime$=Remove(B,NH);} 
                     \State{PR$^\prime$=PR$\times$(1-p$_{NH}$);}
        	        \State{[CP$^\prime$, PR]=HFES(NH,d,R,B$^\prime$,NC$^\prime$,FP,p,PR$^\prime$,CP$^\prime$);}
        	        \If{PR$\neq$ -1}
        	            \State{CP=CP$^\prime$;}
        	        \EndIf
        	    \doWhile{NH$\neq$ $\emptyset$}
        	\EndIf\\
    	    \Return{$<$CP,\:-1$>$}
    	\end{algorithmic}
	\end{algorithm}

    \begin{algorithm}
    	\caption{Pseudo-Code of the non-recursive HFES algorithm}
    	\label{alg:HFESNOFESecursive}
    	\allowdisplaybreaks
    	\begin{algorithmic}[1]
        	\INPUT{CN, d, R, B, P, F, C}
        	\OUTPUT{$<$CP$>$}
        	\For{f $\in \mathcal{F}$}
        	\While{R$\neq \emptyset$}
        	        \State{B$^\prime$=PruneLinks(B,C);}
        	        \State{SP=ShortestPathBasedOnFault(CN,d,B$^\prime$);}
        	        \State{SP=PruneFog Nodes(SP,C);}
        	        \State{$<$NH,pth$>$=SelectNextNodeEnergy-Aware(CN,d,B$^\prime$);}
        	        \State{CP=\{pth\}$\cup$ CP;}
        	        \State{B=Remove(B,pth);}
        	        \State{R=ProvideServices(R,pth,C,F);}
        	\EndWhile
        	\State{B$^\prime$=PruneLinks(B,C);}
        	\State{SP=ShortestPathBasedOnFault(CN,d,B$^\prime$);}
        	\State{CP=\{SP\}$\cup$ CP;}
        	\EndFor\\
    	    \Return{$<$CP$>$}
    	\end{algorithmic}
	\end{algorithm}
    Since the computational complexity of the recursive \textbf{Algorithm~\ref{alg:HFES}} is not deterministic and depends on the input, in order to have an algorithm with a deterministic computational complexity, \textbf{Algorithm~\ref{alg:HFESNOFESecursive}} is proposed which is a greedy non-recursive solution for the proposed optimization problem. In line 1, for each flow the process of resource allocation is done in a sequential manner. Until all of required VNFs are met (line 2) the following actions are repeated:
    \begin{itemize}
    \item all links that have a free capacity less than the required rate of the flow are removed,
    \item based on the fault probability of the switches, shortest paths from the current node $CN$ to all other nodes are calculated in line 4,
    \item Fog Nodes that have a processing capacity less than the required processing capacity for providing service to the flow are removed from the list of shortest paths from $CN$ to other nodes,
    \item at this point, the node $NH$ which has a minimum energy consumption and can provide service to the flow is selected (the shortest path to $NH$  is specified with $pth$) (line 6),
    \item in lines 7 and 8 the shortest path to $NH$ is added to chosen path $CP$ and the passed nodes are removed from bandwidth matrix to prevent loops,
    \item supported VNFs that are requested by the flow are delivered to it in line 9.
    \end{itemize}
    When all of the required VNFs are met by the flow $f$, the flow moves to the destination switch via a path with minimum fault probability.
        

    
    \subsection{Computational Complexity}
	\begin{table*}[!htbp]
    \caption{Traffic Generator Notations and Inputs for 9 scenarios.}\label{tab:TrafficGeneratorNotation}
    \centering
    \scriptsize
    \rowcolors{2}{gray!25}{white}
    \begin{tabular}[t]{|p{0.05\textwidth}|p{0.33\textwidth}||p{0.033\textwidth}|p{0.033\textwidth}|p{0.033\textwidth}|p{0.033\textwidth}|p{0.033\textwidth}|p{0.033\textwidth}|p{0.033\textwidth}|p{0.033\textwidth}|p{0.033\textwidth}|}
    	\hline
    	\textbf{Notation} & \textbf{Definition} & \textbf{S1} & \textbf{S2} & \textbf{S3} & \textbf{S4} & \textbf{S5}& \textbf{S6} & \textbf{S7} & \textbf{S8} &\textbf{S9}\\
    	\hline\hline
    	    $B^f$ & Ratio of flow size to link capacity & 0.01 & 0.05 & 0.1 & 0.05 & 0.05 & 0.05 & 0.05 & 0.05 & 0.05\\\hline 
    	    $\gamma$ & Ratio of Fog Nodes to switches& 0.5 & 0.5 & 0.5 & 0.5 & 0.7 & 1 & 0.5 & 0.5 & 0.5\\\hline 
    	    $R^f$ & Average number of requested VNFs that a flow needs & 2 & 2 & 2 & 2 & 2 & 2 & 2 & 4 & 6\\\hline 
    	    $X_\gamma$ & Ratio of VNFs hosted by a Fog Node & 0.7 & 0.7 & 0.7 & 0.7 & 0.7  & 0.7 & 0.7 & 0.7 & 0.7\\\hline            		    
    	    $R^f_{min}$ & Minimum number of requested VNFs per flow & 2 & 2 & 2 & 2 & 2 & 2 & 2 & 2 & 2\\\hline 
    	    $R^f_{max}$ & Maximum number of requested VNFs per flow & 5 & 5 & 5 & 5 & 5 & 5 & 5 & 5 & 5\\\hline            		    
    	    $\tau$ & Edge switches ratio & 1 & 1 & 1 & 1 & 1 & 1 & 1 & 1 & 1\\\hline 
    	    $\tau_s$ & Ratio of edge switches that are source of a flow & 1 & 1 & 1 & 1 & 1 & 1 & 1 & 1 & 1\\\hline
    	    $\tau_d$ & Ratio of edge switches that are destination of a flow & 1 & 1 & 1 & 1 & 1 & 1 & 1 & 1 & 1\\\hline 
            $\omega$ & Coefficient of number of generated flows per source & 0.4 & 0.4 & 0.4 & 0.4 & 0.4 & 0.4 & 0.4 & 0.4 & 0.4\\\hline 
            $F_m$ & Maximum number of generated flows per source & 10 & 10 & 10 & 10 & 10 & 10 & 10 & 10 & 10\\\hline
    	    $X$ & Number of different VNFs & 10 & 10 & 10 & 10 & 10 & 10 & 10 & 10 & 10\\\hline
    \end{tabular}
    \end{table*}
    \noindent\textbf{OFES:} The problem can be reduced to capacity-aware multi commodity problem which is categorized as an NP-hard problem.
        
    \noindent\textbf{HFES:} The computational complexity of lines 1 and 2 of algorithm \ref{alg:HFESNOFESecursive} are in the order of $O(F)$ and $O(\psi)$, respectively. The order of the computational complexity of lines 3, 4, 5, and 12 are $O(E+N\cdot \log{N})$ while it is $O(N)$ for lines 6, 7, 8, and 9. Similarly, the computational complexity of line 11 is $O(E)$. Therefore, the total computational complexity of \textit{HFES} is $O\left(F\cdot \psi\cdot \left[N\cdot \log{N}+E+N\right]\right)$. 
        
\section{Scenario Description}\label{sec:scenariodescription}
The following subsections detail the pursued scenarios, methodology and the traffic generator for the applied scenarios. 
    \subsection{Simulation Setup}
\begin{figure}[!htbp]
	    \centering
		\includegraphics[width=0.6\columnwidth]{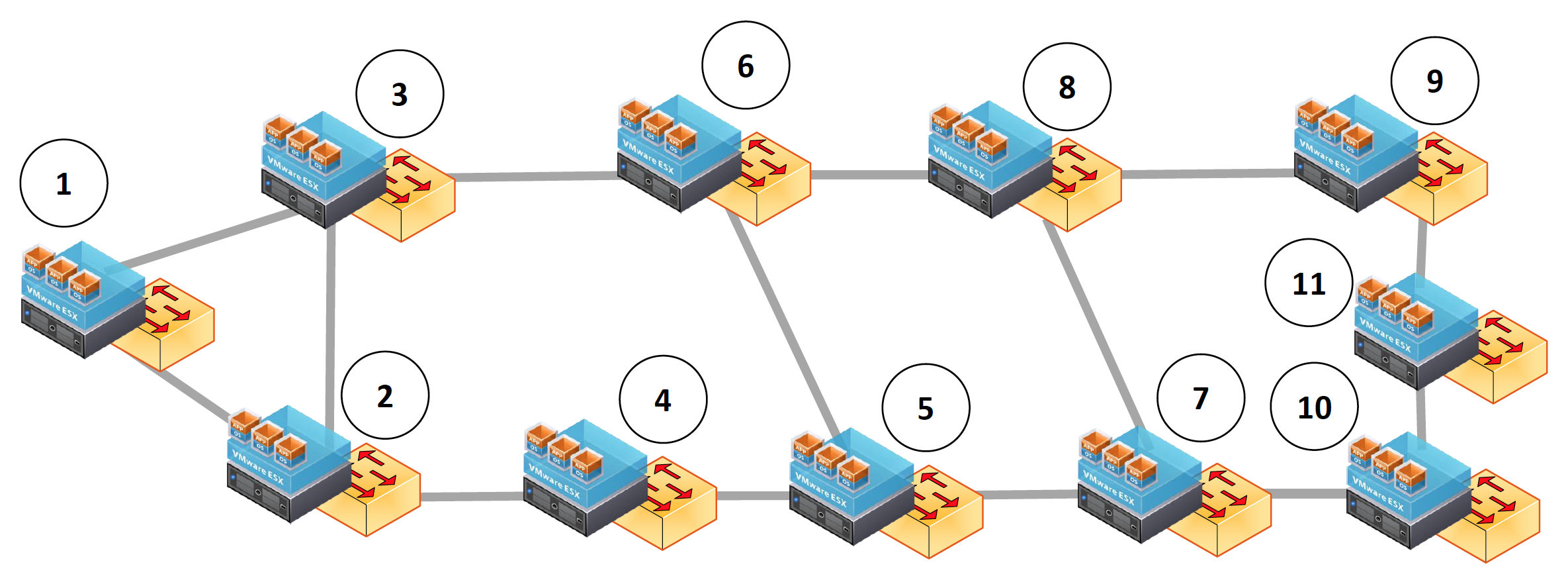}
    	\caption{Abilene Network Topology.}
    	\label{fig:abileneTopology}
    \end{figure}

    In this paper, we consider Abiliene~\cite{AbilineNetwork} as the network topology which is shown in Fig \ref{fig:abileneTopology}.
 We consider the capacity of all links (maximum bandwidth) to be equal to $1$ Gbps. Each Fog Node has two states: ON (i.e., working with full rate energy consumption) and SLEEP (i.e., working with a fix minimum energy consumption). The processing power of a Fog Node is a factor of the input bandwidth. Similarly, the energy consumption is a factor of processing power. Traffic flows are generated using geometric distribution where the rate of flows are a fraction of links bandwidth, referred to $B^f$. 


\subsection{Traffic-Demand Generator}\label{sec:trafficGenerator}
	In order to investigate the performance of these resource allocator/reallocator algorithms, a traffic-demand generator is proposed. It takes multiple input parameters and generates network traffic flows with different specifications: rate, source and destination, VNF requirements, and end-to-end tolerable delay. Table~\ref{tab:TrafficGeneratorNotation} presents the input parameters of the traffic generator. It is important to note that the described algorithm is meant to generates the traffic pattern (and not the traffic packets).  
    We consider the flows as unidirectional. The variable $\tau$ specifies the percentage of switches that act as edge switches. In other words, variable $\tau$ stands for the percentage of switches that can be source or destination of a flow. As an example, $\tau=1$ means that all switches can be considered as an edge switch while $\tau=0$ means that there is no edge switch. Therefore, the number of edge switches is $\tau\times N$. Similarly, the variables $\tau_s$ and $\tau_d$ are the percentage of edge switches that can act as the source or the destination of a flow, respectively. Thus, the number of source switches is $N_s = \tau\times\tau_s\times N$ and the number of destination switches is $N_d = \tau\times\tau_d\times N$. The number of possible source-destination pairs is $N_s\times N_d$.
        
    We want $\omega\times N_d$ to be the average number of flows that are generated by a source node. We assume that the number of generated flows for each source switch follows a geometric distribution \cite{zoucomputer} with $1/(\omega\times N_d)$ as the success probability. For practical reasons, we also include in the model the maximum number of flows $F_m$ that can be generated by a source node. Therefore, the flows will be generated with a truncated geometric distribution and the average number of flows from each source node will be smaller than $\omega\times N_d$. $\gamma$ is the ratio of Fog Nodes to switches, therefore, $\gamma\times N$ is the number of Fog Nodes. On the other hand, $X_\gamma$ is the fraction of VNF types hosted by a Fog Node, meaning that a Fog Node can host at most $X_\gamma\times X$ different VNFs, where $X$ is the number of different types of VNFs. The number of VNFs that are needed by a flow is generated according to geometric distribution with average $R^f$. In order to make realistic scenarios, we consider $R^f_{max}$ and $R^f_{min}$ as the maximum and minimum number of VNFs that a flow needs, respectively. If the generated number is greater than $R^f_{max}$, the number is set to $R^f_{max}$ and similarly for the minimum (note that this approach changes the average number with respect to the average of the initial geometric distribution). The average traffic rate demand of a flow is a fraction $B^f$ of the capacity of the link, i.e., it is $B^f\times link\_capacity$. In particular, the rate of generated flows follows a uniform distribution between 0 and $2\times B^f\times link\_capacity$. According to the Fig. \ref{fig:abileneTopology}, the number of switches, links and functions (or VNFs) are 11, 14, and 10, respectively. Moreover, Maximum joint failure probability equal to 0.1, link propagation delay 100 [ms], and $TP_x=3$ [ms].
    
    \subsection{Scenarios}
    In order to investigate the impact of the different traffic patterns and network resources we evaluate the performance of the proposed solutions over different traffic scenarios which are presented in Table \ref{tab:TrafficGeneratorNotation}. We generate the traffic demands based on three main characteristics: i) flow size, ii) number of Fog Nodes, and, iii) number of required VNFs. In order to evaluate the impact of the flow size, three different values for $B^f$ (i.e., 0.01, 0.05 and 0.1\}) are considered (see the three first scenarios, S1, S2 and S3, in Table \ref{tab:TrafficGeneratorNotation}). By changing the value of $\gamma$ among \{0.5, 0.7, or 1\} the impact of the number of Fog Nodes is investigated (see the scenarios 7 to 9). Finally, to investigate the impact of number of VNFs required by the flows, three different values for $R^f$ (i.e., 2, 4, and 6) are considered (see the last three scenarios).
    
\section{Simulation Results} \label{sec:results}
    In this section, proposed solutions named OFES and HFES are compared using several metrics (see subsections~\ref{subsec:EnFpPlNs}). In this section, the solutions are compared over i) total energy consumption of the processing Fog Nodes, ii) average fault probability of the selected paths, iii) average path length, and, iv) the side-effect of network reconfiguration.
    
    \subsection{Energy Consumption}\label{sec:8.1}
\begin{figure}[!htbp]
	    \centering
		\includegraphics[width=0.33\linewidth]{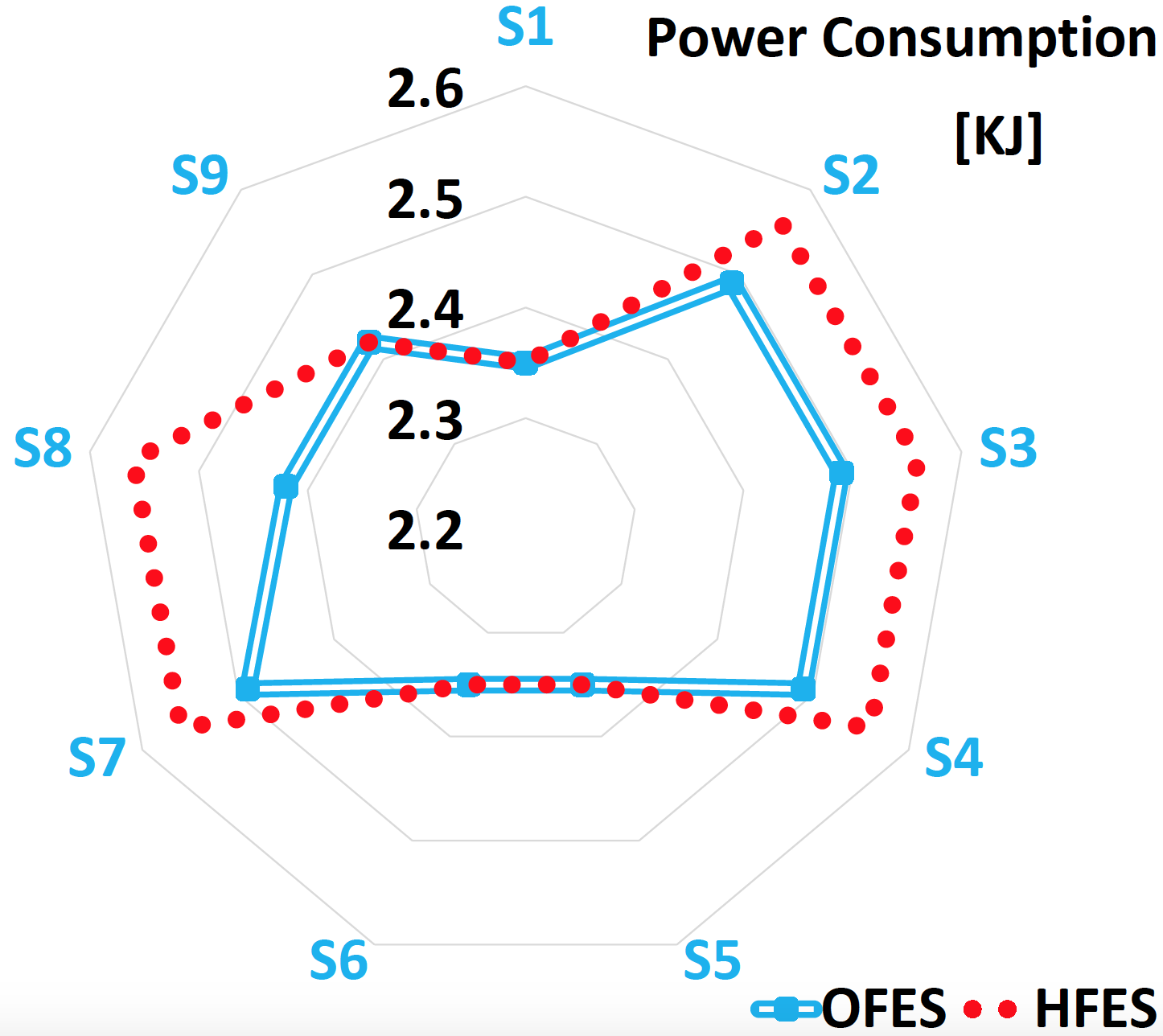}
    	\caption{Power Consumption.}
    	\label{fig:spdPW}
    \end{figure}

    Figure~\ref{fig:spdPW} compares the energy consumption of OFES and HFES in different traffic scenarios. In this figure, the blue and red points in $Si$ specifies the energy consumption of OFES and HFES in traffic scenario $i$, respectively. From Fig.~\ref{fig:spdPW}, the energy consumption of HFES is near the energy consumption of the OFES (which is the optimal solution) in all traffic scenarios. In order to investigate the impact of increasing the average rate of flows, the value of red and blue points in $S1$, $S2$, and $S3$ should be compared. Increasing the average rate of flows increases the total energy consumption in both OFES and HFES. Considering the difference between the result of OFES and HFES as the optimality gap of HFES, increasing the average rate of flows increases the optimality gap. On the other hand, increasing the number of Fog Nodes ($S4$, $S5$, and $S6$) decreases both the total energy consumption and the optimality gap. However, increasing the average number of required VNFs per flow ($S7$, $S8$, and $S9$) does not have a predicable impact on the optimality gap. Our simulations show that the energy consumption of HFES is at most 3\% more than the energy consumption of OFES.
    
    \subsection{Average Fault Probability}\label{sec:8.2}
    \begin{figure}
	    \centering
		\includegraphics[width=.33\textwidth]{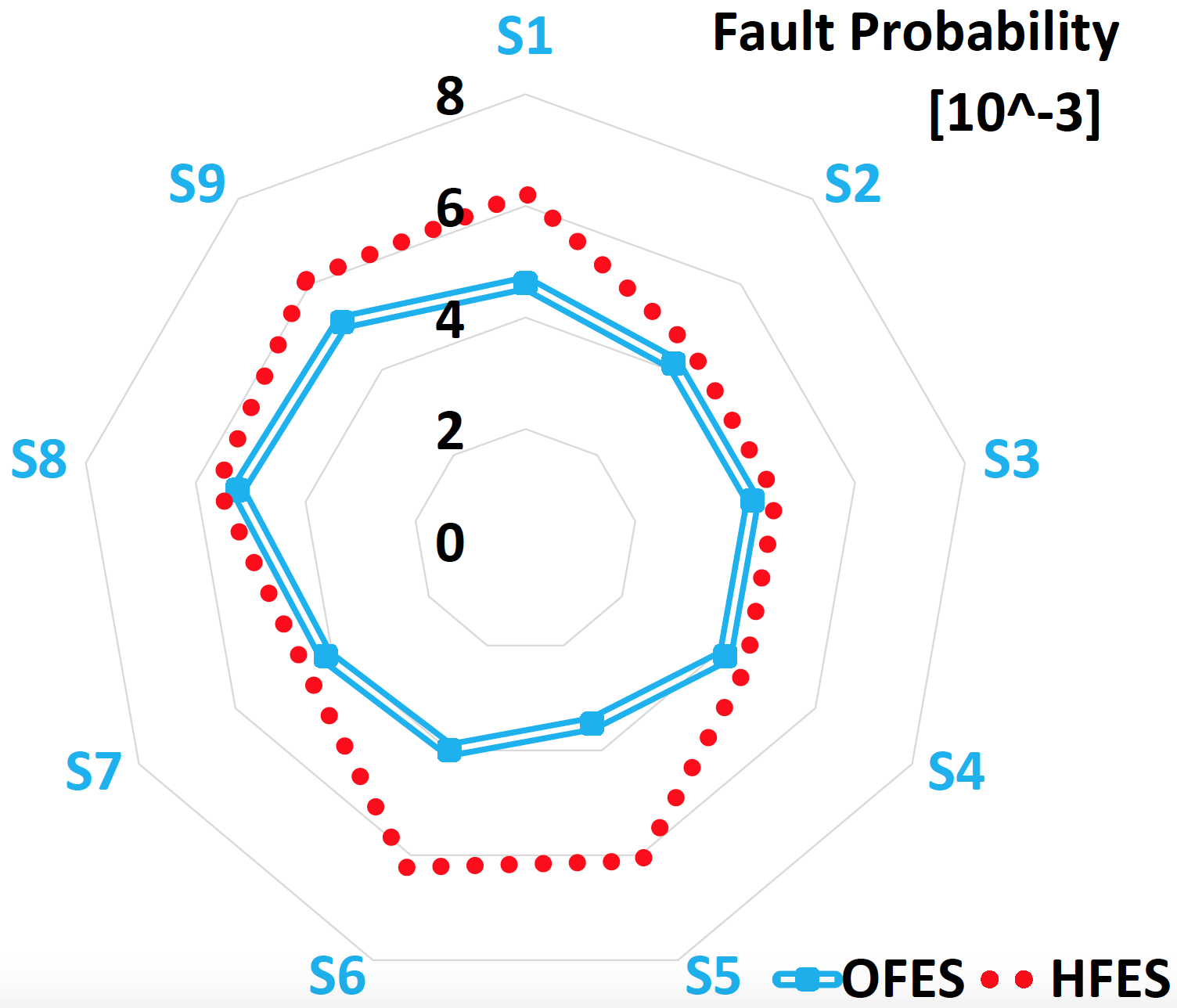}
    	\caption{Fault Probability.}
    	\label{fig:spdFP}
	\end{figure}
    Figure~\ref{fig:spdFP} presents average fault probability of selected paths of OFES and HFES in different traffic scenarios. Based on this figure, increasing average flow rate decreases the average fault probability of both OFES and HFES. This happens because increasing the average rate of flows forces the algorithms to turn ON more Fog Nodes to serve the flows. This helps the algorithms to find low faulty paths for flows. In brief, increasing the rate of flows increases the number of active Fog Nodes and the energy consumption but decreases the fault probability. On the other hand, it decreases the optimality gap of fault probability of HFES because the number of different paths that a flow can go through decreases (due to links capacity limit). 
    Increasing the number of Fog Nodes increases the optimality gap of HFES since it increases the number of links that flows can cross and consequently increases the number of possible valid results. When the number of valid results increases the probability of HFES on finding a non-optimal result increases. Similar to the previous subsection, the impact of increasing the number of required VNFs per flow on the average fault probability is not predictable.
    
    \subsection{Average Path length and Network Side-effect} \label{subsec:EnFpPlNs}
\begin{figure}[!htbp]
    \centering
    	\begin{subfigure}{0.49\textwidth}
    	    \centering
    		\includegraphics[width=0.65\linewidth]{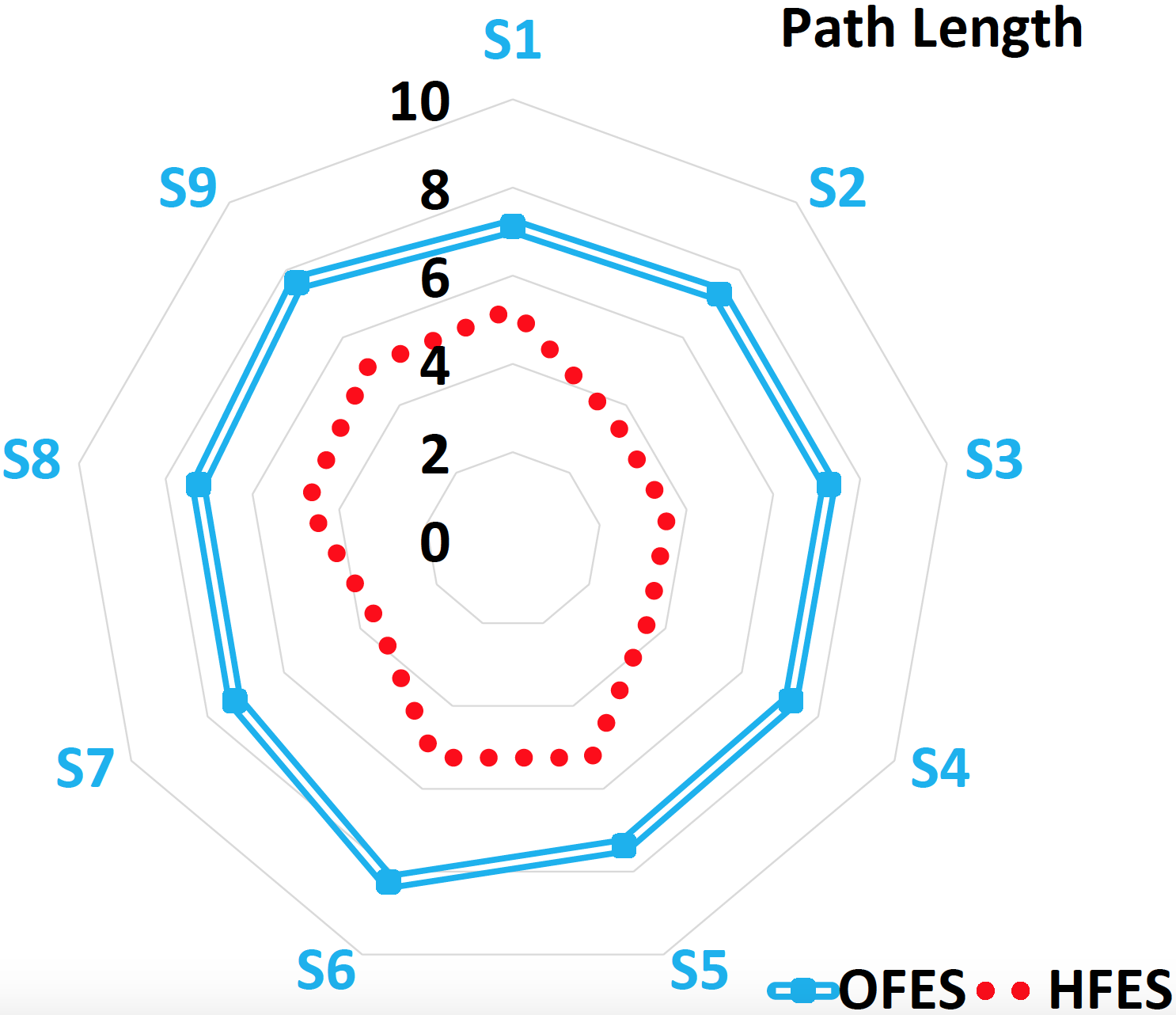}
        	\caption{Path Length.}
        	\label{fig:spdPL}
    	\end{subfigure}
    	\begin{subfigure}{0.49\textwidth}
    	    \centering
    		\includegraphics[width=0.65\linewidth]{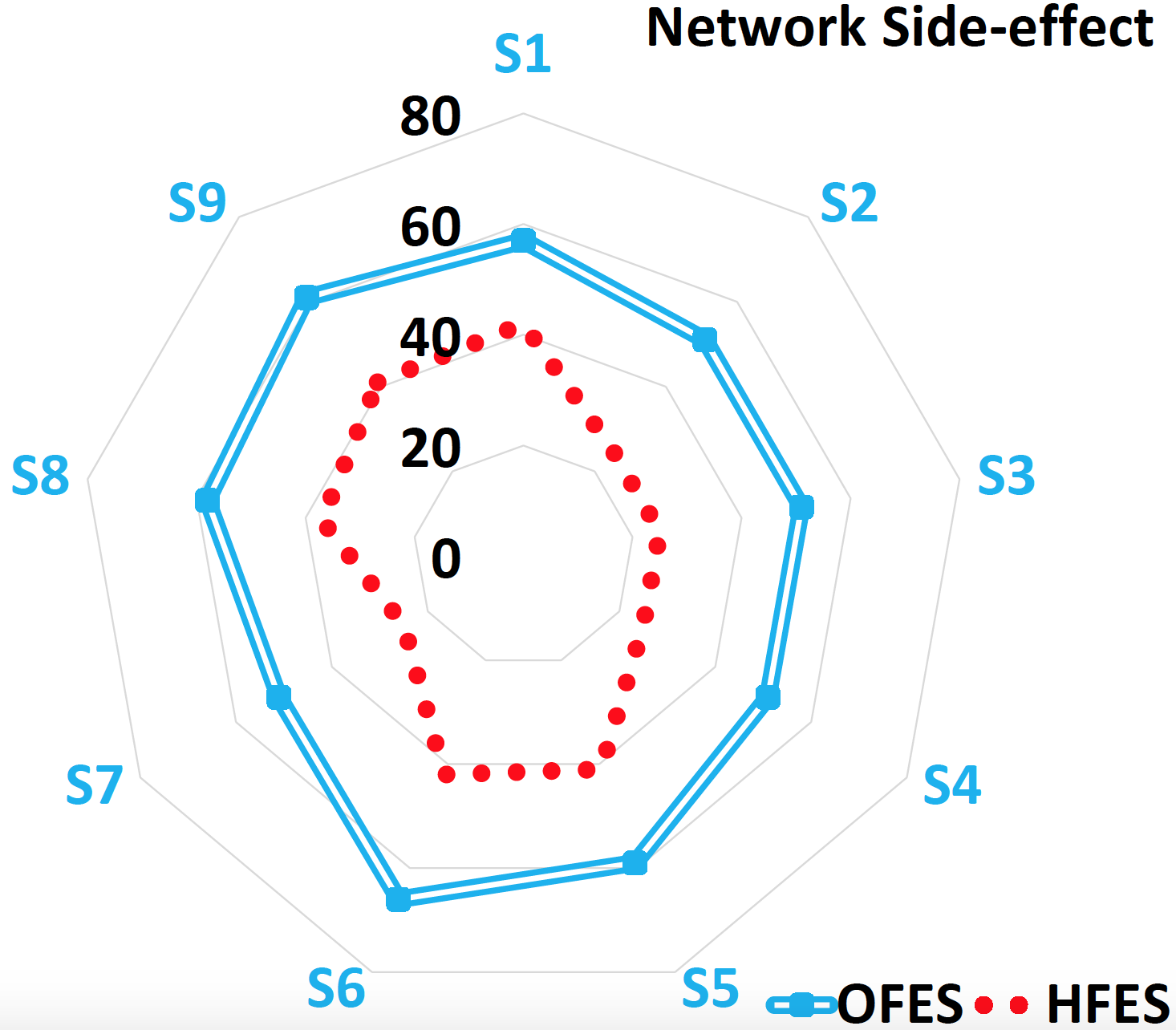}
        	\caption{Network Side-effect.}
        	\label{fig:spdNS}
    	\end{subfigure}
    	\caption{ \textit{OFES} vs. \textit{HFES} Comparisons.}
        \label{fig:spdPLNS}
    \end{figure}

    We consider the number of forwarding table entries that need to be setup in network reconfiguration as network side-effect of a solution.
    Figure~\ref{fig:spdPLNS} shows the average path length and network side-effect of failure recovery using HFES and OFES in different traffic scenarios. The figure shows that the average path length and network side-effect of reconfiguration in HFES are lower than OFES. This is because the focus of OFES is on optimization of power consumption and the fault probability. HFES can reduce the average path length and network side-effect up to 50\% compared with OFES.
    
    \subsection{Link and Fog Node Utilization}
\begin{figure}[!htbp]
    \centering
        \begin{subfigure}{0.49\columnwidth}
    	    \centering
    		\includegraphics[width=0.7\linewidth]{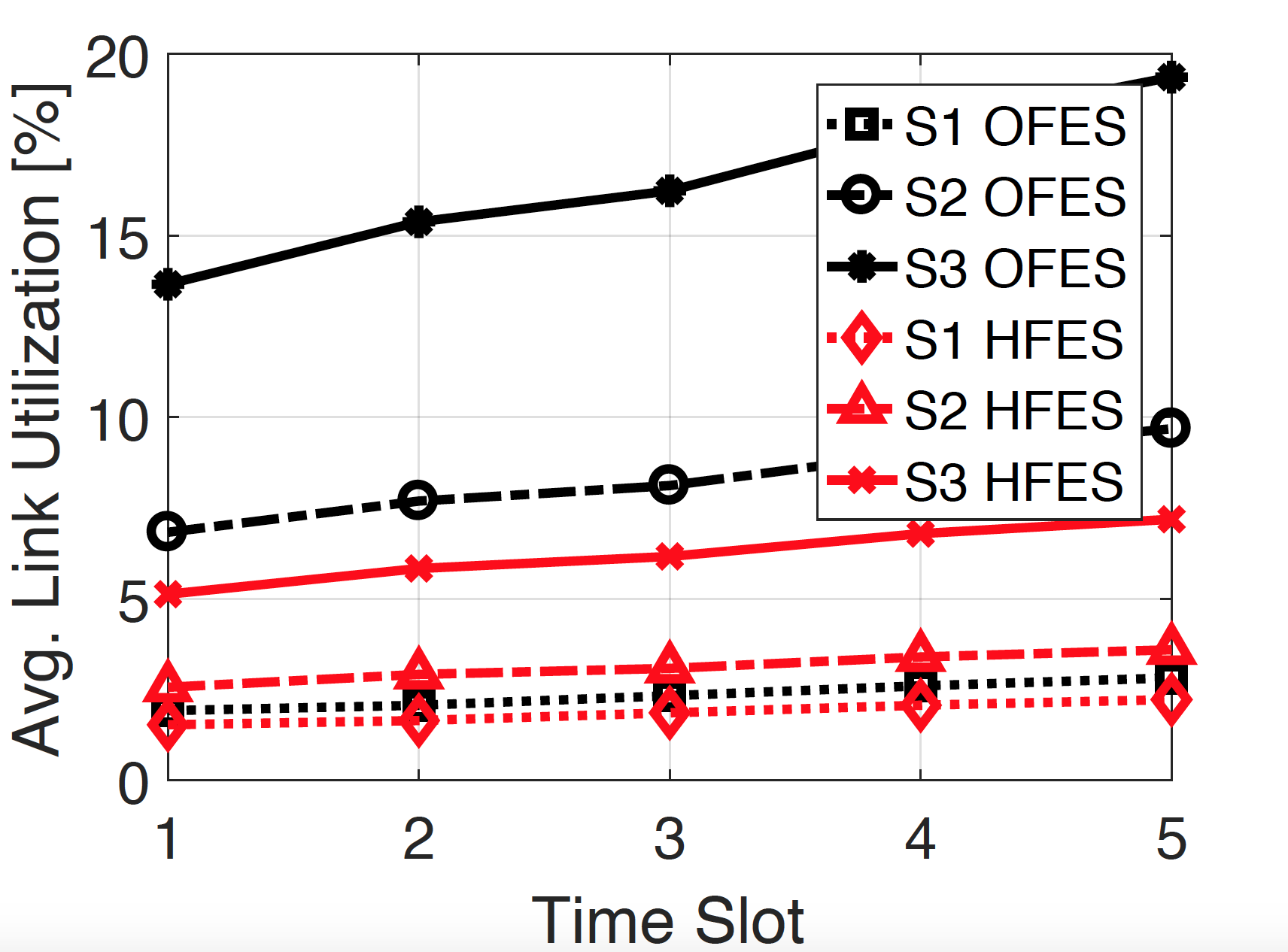}
        	\caption{Link Utilization.}
        	\label{fig:LUS123}
    	\end{subfigure}
    	\begin{subfigure}{0.49\columnwidth}
    	    \centering
    		\includegraphics[width=0.7\linewidth]{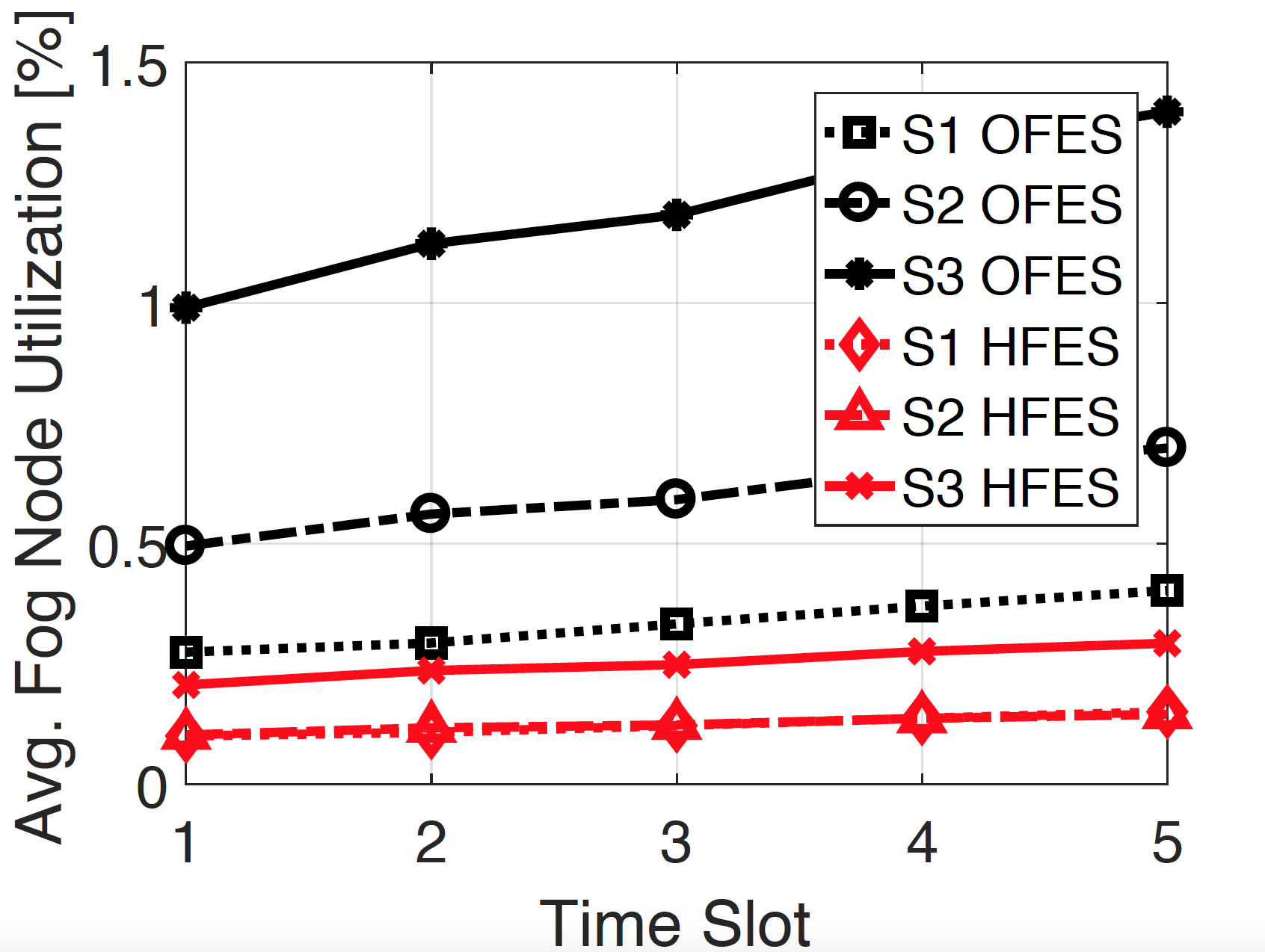}
        	\caption{Fog Node Utilization.}
        	\label{fig:NUS123}
    	\end{subfigure}
    	\caption{Impact of Flow Rate.}
        \label{fig:NodeLinkUtilizationS123}
    \end{figure}

    In this subsection, we investigate the impact of the flow rate, number of Fog Nodes, and number of required VNFs on link and Fog Node utilization. Since the focus of the OFES is on optimizing the energy consumption and the probability of fault, the average link and Fog Node utilization of HFES is lower than OFES.
    Figure~\ref{fig:NodeLinkUtilizationS123} shows the impact of average flows rate on link and Fog Node utilization. As can be seen, increasing the average flows rate increases the average link and Fog Node utilization. Increasing the flows rate lead to more higher traffic load inputs, therefore, the link utilization increases. Similarly, the amount of data that should be processed by the Fog Nodes increases and consequently the Fog Nodes utilization grow up.

\begin{figure}
    \centering
    	\begin{subfigure}{0.49\columnwidth}
    	    \centering
    		\includegraphics[width=0.7\linewidth]{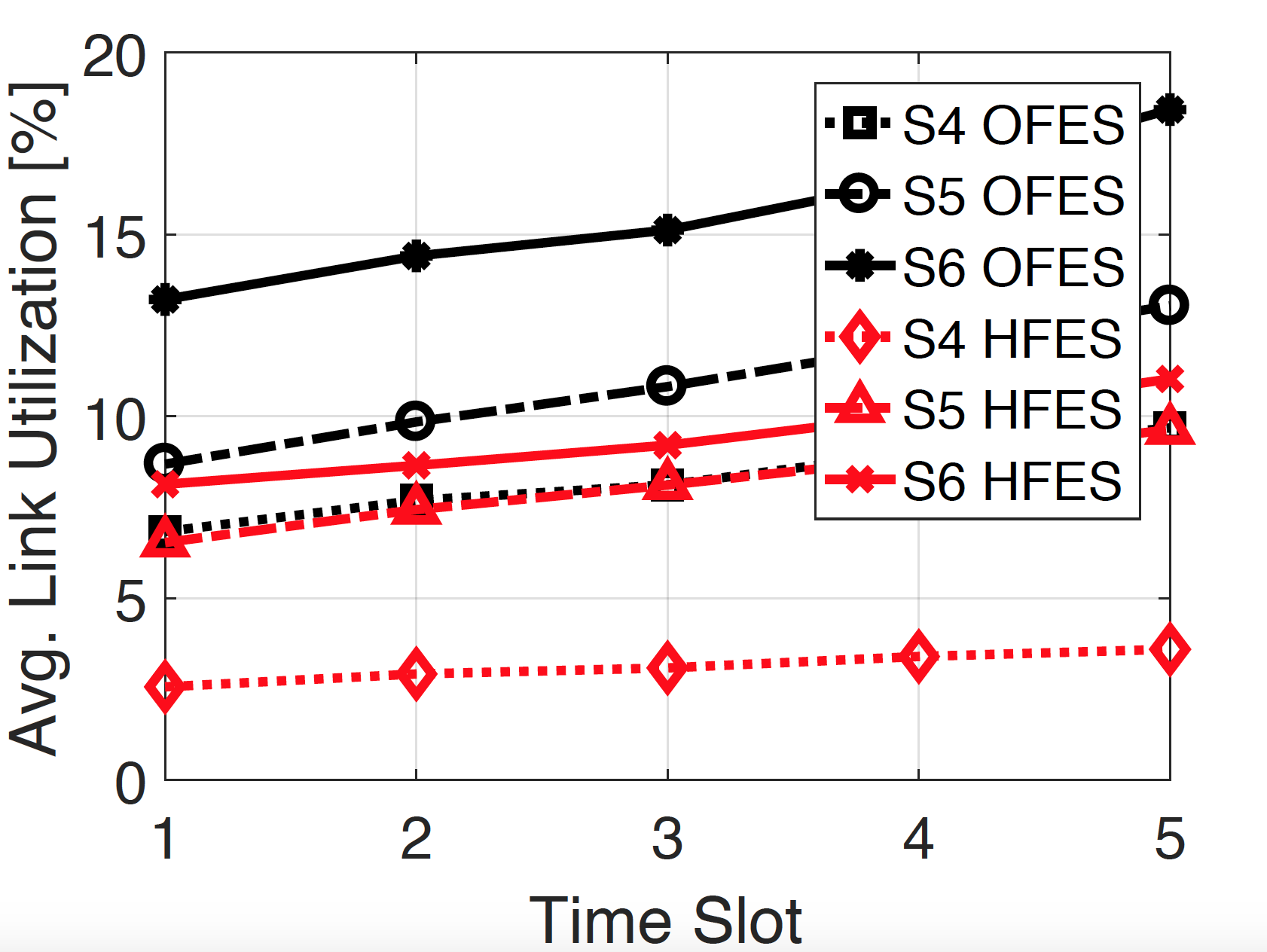}
        	\caption{Link Utilization.}
        	\label{fig:LUS456}
    	\end{subfigure}
    	\begin{subfigure}{0.49\columnwidth}
    	    \centering
    		\includegraphics[width=0.7\linewidth]{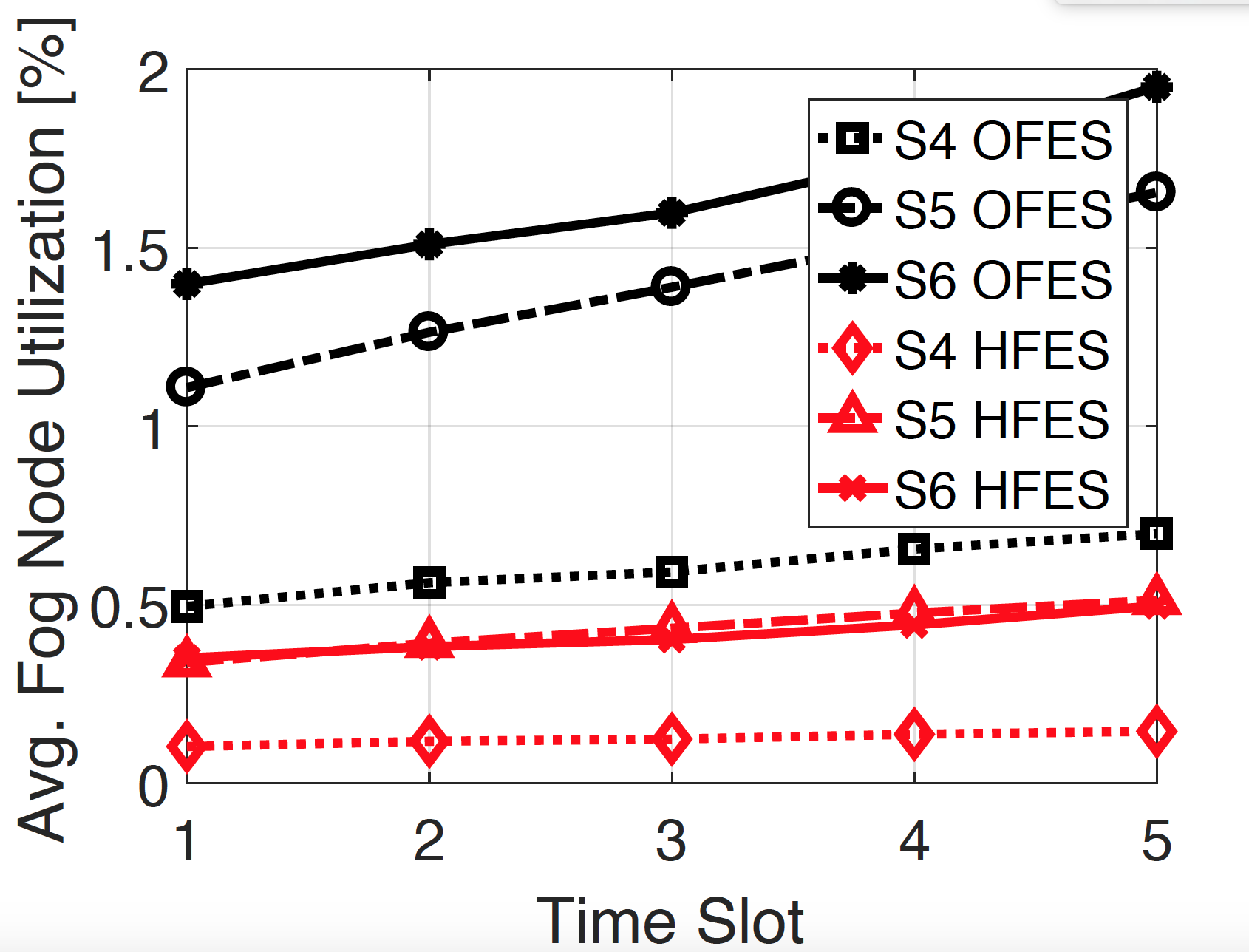}
        	\caption{Fog Node Utilization.}
        	\label{fig:NU2}
    	\end{subfigure}
    	\caption{Impact of number of Fog Nodes.}
        \label{fig:NodeLinkUtilizationS456}
    \end{figure}

    From Fig.~\ref{fig:NodeLinkUtilizationS456} one can say that increasing the number of Fog Nodes increases the link utilization and Fog Node utilization. This happens because on one hand, when the number of Fog Nodes are increased, both HFES and OFES have a higher chance to find a hub Fog Node which has enough processing power to serve a lots of flows. On the other hand, the algorithms tries to reduce the energy consumption, therefore, they sends the unnecessary Fog Nodes to the IDLE mode. This leads to an increment in the Fog Nodes utilization since some nodes are not working and consequently they are not considered in the utilization measurement (Fig.~\ref{fig:LUS456}. Similarly, since the algorithms tries to bring lots of flows into hub Fog Nodes, the path length and link utilization increases (Fig.~\ref{fig:NU2}.
    
\begin{figure}[!htbp]
    \centering
    	\begin{subfigure}{0.49\columnwidth}
    	    \centering
    		\includegraphics[width=0.7\linewidth]{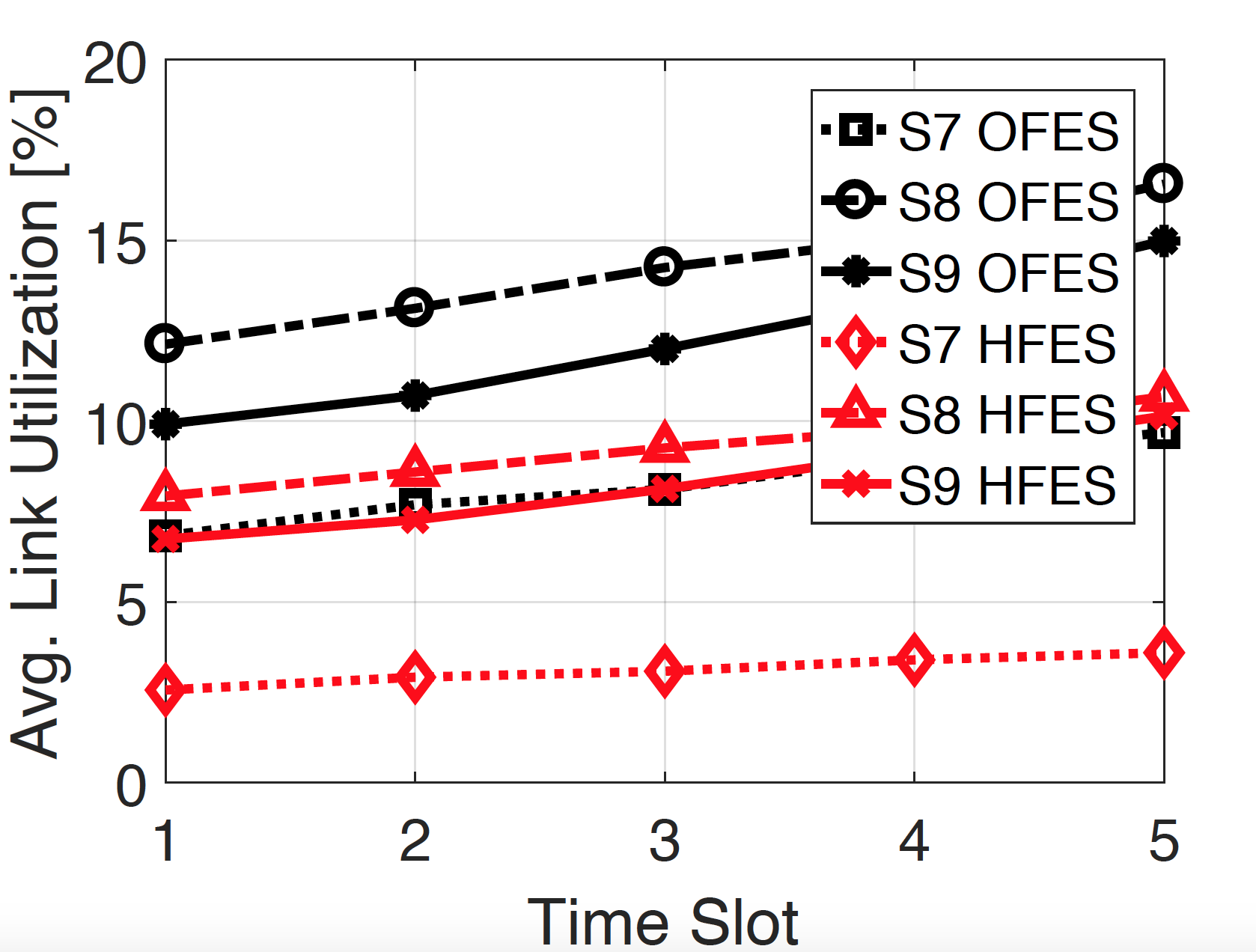}
        	\caption{Link Utilization.}
        	\label{fig:LUS789}
    	\end{subfigure}
    	\begin{subfigure}{0.49\columnwidth}
    	    \centering
    		\includegraphics[width=0.7\linewidth]{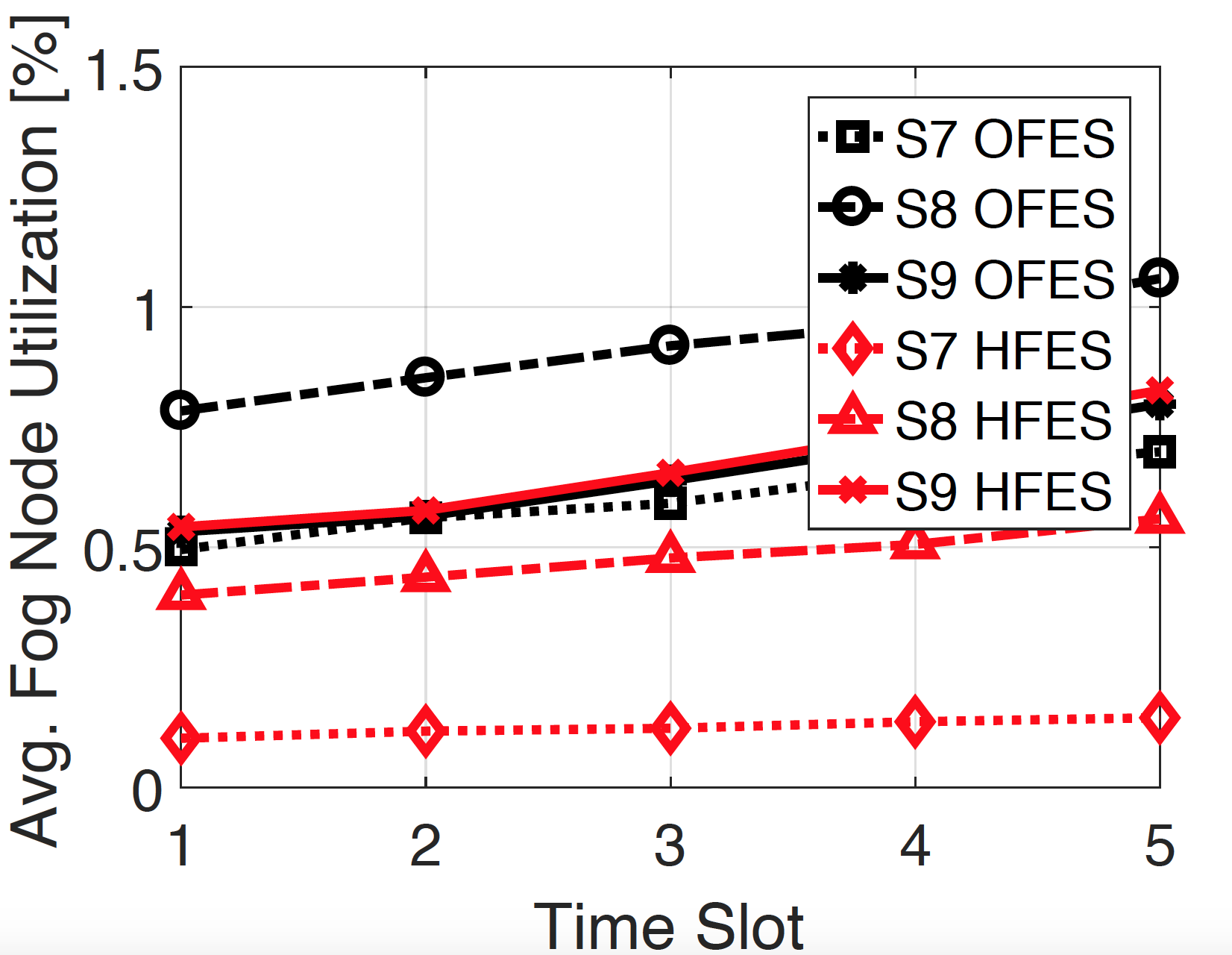}
        	\caption{Fog Node Utilization.}
        	\label{fig:NUS789}
    	\end{subfigure}
    	\caption{Impact of Number of Required VNFs per flow.}
        \label{fig:NodeLinkUtilizationS789}
    \end{figure}

    Figure~\ref{fig:NodeLinkUtilizationS789} investigate the impact of number of required VNFs per flow on the link and Fog Node utilization. As can be seen, the number of required VNFs does not have a predicable impact on these metrics. This is because when the number of required VNFs increases, the algorithms try to provides service to the flows without using new Fog Nodes, therefore, the average link and Fog Node utilization increases. But when the requests go high, the algorithms turn new Fog Nodes ON, therefore, the average link and Fog Node utilization decreases.
    
\begin{figure}[!htbp]
	    \centering
		\includegraphics[width=0.75\textwidth]{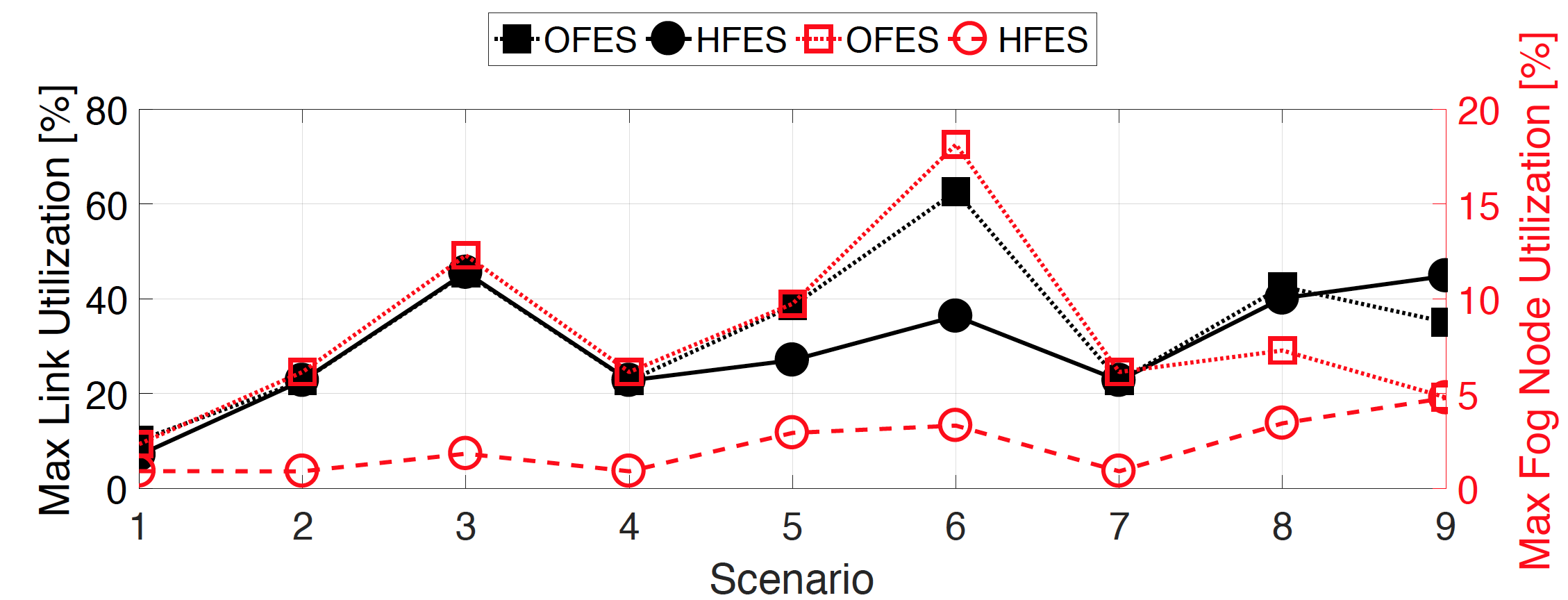}
    	\caption{Max Link and Fog Node Utilization.}
    	\label{fig:MaxUtilization}
	\end{figure}

    Figure~\ref{fig:MaxUtilization} presents the maximum link and Fog Node utilization versus different scenarios. As can be seen, increasing average flows rate increases the maximum link and Fog Node utilization. It has a higher impact on the maximum link utilization compared with maximum Fog Node utilization. Similarly, increasing the number of Fog Nodes increases the maximum link utilization of OFES dramatically more than the maximum link utilization of HFES. This happens because when the number of Fog Nodes increases the chance of OFES for finding a hub Fog Node and uses that Fog Node to serve higher number of flows increases. Therefore, the maximum link utilization of the network grows because more flows crosses the hub Fog Node. Additionally, when a hub Fog Node is detected OFES puts a higher number of Fog Nodes in IDLE mode, therefore, the maximum link and Fog Node utilization increases. Finally the impact of increasing the number of required VNF on the OFES is not predicable while it increases the maximum link and Fog Node utilization in HFES. When the number of required VNFs increases, OFES turns ON more Fog Nodes to deliver service to the flows, therefore, the maximum link and Fog Node utilization decreases in scenario 9 compared to scenario 8. On the other hand, when the number of required VNFs increases if the algorithms do not put new Fog nodes in ON mode, the maximum Fog Node utilization increases since the amount of request is increased. Similarly, the flows have to pass extra hops to receive all of required VNFs, therefore, the maximum link utilization increases, too.

\subsection{Impact of $\alpha$ and $\beta$ on OFES}
    In order to investigate the impact of $\alpha$ and $\beta$ in Eq.~\eqref{eqObjective} on the performance of the optimal solution, the performance of OFES for different values of $\alpha$ and $\beta$ is presented in Table~\ref{tab:alphabetaeffectonObjective}. To this end, we set $\beta=1-\alpha$ and measure the energy consumption and the network side-effect of OFES. As expected, increasing the value of $\alpha$ increases the network side-effect but decreases the energy consumption.
   	\begin{table}[!htbp]
        \caption{Impact of $\alpha$ and $\beta$ in OFES.}\label{tab:alphabetaeffectonObjective}
        \centering
        \begin{tabular}[t]
        {|l|l|l|l|l|l|l|l|}
            \hline
            \multirow{2}{*}{\textbf{Methods}}&\multicolumn{7}{c|}{$\alpha$}\\\cline{2-8}
            &\textbf{0} & 0.001 & 0.004 & 0.005 & 0.1 & 0.75 & 1\\\hline
           \coldscr\textbf{$E(t)[kJ]$}&\coldscr 1.6&\coldscr 1.2&\coldscr 1.2&\coldscr 0.8&\coldscr 0.8&\coldscr 0.8&\coldscr 0.8\\\hline
           \textbf{$NS$}&19&19&19&21&21&21&53\\\hline
        \end{tabular}
    \end{table}
    
\section{Conclusion and Future Work}\label{sec:conclusions}
In this paper, an efficient failure recovery and fault prevention algorithm for SDN-based networks was introduced. The problem was mathematically formulated and an optimal scheme was proposed to solve the corresponding optimization problem called OFES. The proposed formulation optimizes the Fog Nodes' energy consumption while guarantying the QoS constraints. Due to high computational complexity of the proposed solution, we introduced a heuristic approach, called HFES, which is a sub-optimal solution with low computational complexity. The computational complexity of HFES was discussed and showed that it is applicable to real-world networks. HFES was compared with OFES for power consumption, fault probability, average path length, network side-effect, and average link and node utilization. Additionally, the impact of flow rate, number of Fog Nodes, and the number of required VNFs on the proposed algorithms was discussed. Our simulations show that the energy consumption of HFES is at most 3\% more than OFES, while the average path length and network side-effect of HFES is 50\% less than OFES. Besides, the maximum link and Fog Node utilization of HFES are up to 75\% less than those of the OFES. Both OFES and HFES can assign resources in a way that the average fault probability stays below a predefined threshold, however, the average fault probability of OFES is at most 40\% lower than HFES. Due to high computational complexity of OFES, its solution is applicable only for very small networks, while HFES are applicable to large-scale networks. Future works will be dedicated to consider the energy consumption of networking devices (switches and links). To this end, switches and links that are not used should be turned off. Additionally, the formulation and the heuristic algorithm could be extended to consider the VNFs ordering. Another field of interest is consideration of queuing delay in configuration of the network.

\section*{Acknowledgment}\label{sec:10}
This work has received funding from the Horizon 2020 EU project SUPERFLUIDITY (grant agreement No. 671566). Mauro Conti is supported by a Marie Curie Fellowship funded by the European Commission (agreement PCIG11-GA-2012-321980). This work is also partially supported by the EU TagItSmart! Project (agreement H2020-ICT30-2015-688061), the EU-India REACH Project (agreement ICI+/2014/342-896), by the project CNR-MOST/Taiwan 2016-17 ``Verifiable Data Structure Streaming", the grant n. 2017-166478 (3696) from Cisco University Research Program Fund and Silicon Valley Community Foundation, and by the grant "Scalable IoT Management and Key security aspects in 5G systems" from Intel.

\bibliographystyle{elsarticle-num}

\bibliography{references}

\newpage
\section*{Biographies}\label{sec:11}

\noindent\begin{minipage}{0.1\textwidth}
\includegraphics[width=1.15in,height=1.15in,clip,keepaspectratio]{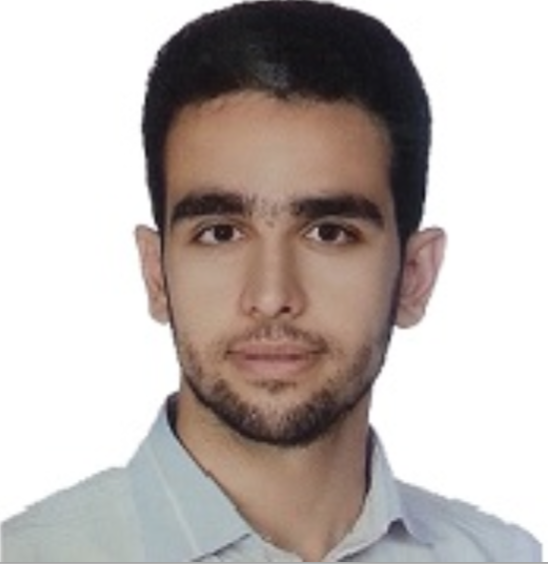}
\end{minipage}%
\hfill%
\begin{minipage}{0.82\textwidth}
\textbf{Mohammad M. Tajiki}
is a PhD candidate at Tarbiat Modares University, spending his sabbatical period in University of Rome Tor Vergata. His main research interests are Network QoS, media streaming over the Internet, data center networking, traffic engineering, service function chaining, IPv6 segment routing, and software-defined networking (SDN).
\end{minipage}%

\hfill \break

\noindent\begin{minipage}{0.1\textwidth}
\includegraphics[width=1.15in,height=1.15in,clip,keepaspectratio]{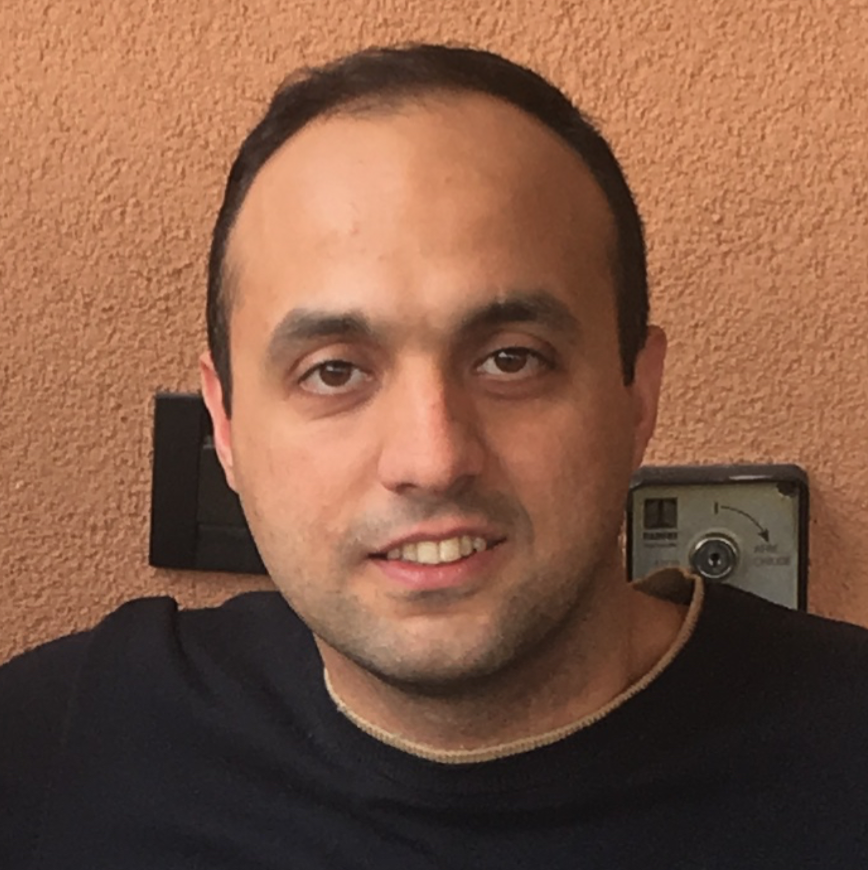} 
\end{minipage}%
\hfill%
\begin{minipage}{0.82\textwidth}
\textbf{Mohammad Shojafar} is currently an Intel Innovator and Senior Researcher in SPRITZ Security and Privacy Research Group at the University of Padua, Italy. He was CNIT Senior Researcher at the University of Rome Tor Vergata contributed on European Horizon 2020 ``SUPERFLUIDITY’’ project. Also, he contributed in some Italian projects named ``SAMMClouds'', ``V-FoG'' and ``PRIN15''  which are supported by the University of Sapienza Rome and the University of Modena and Reggio Emilia, Italy, respectively. He received the Ph.D. degree in ICT from Sapienza University of Rome, Rome, Italy, in May 2016. He is an author/co-author of 88+ peer-reviewed publications (h-index=20, citations=1436+) in prestigious conferences (e.g., ICC, GLOBECOM, ISCC, MASS) and journals in IEEE, Elsevier, and Springer publishers. He served as an associate editor in Springer Cluster Computing and an editor in TJCA and WMWN as TPC in several conferences such as I-SPAN, ICWMC, and UCC. Since 2013, he is the membership of IEEE Systems Man and Cybernetics Society Technical Committee on Soft Computing. His research interests include 5G networks,  cloud data centers,  network security, and optimization techniques. He was a programmer and analyzer in exploration directorate section at National Iranian Oil Company (NIOC) and Tidewater Co. in Iran from 2008-2013, respectively.

\end{minipage}%

\hfill \break

\noindent\begin{minipage}{0.1\textwidth}
\includegraphics[width=1.15in,height=1.15in,clip,keepaspectratio]{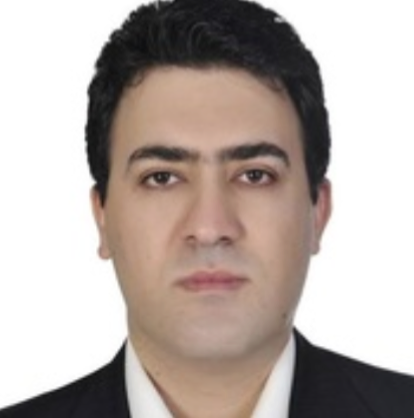} 
\end{minipage}%
\hfill%
\begin{minipage}{0.82\textwidth}
\textbf{Behzad Akbari} received the B.S., M.S., and PhD degree in computer engineering from the Sharif University of Technology, Tehran, Iran, in 1999, 2002, and 2008 respectively. His research interest includes Computer Networks, Multimedia Networking Overlay and Peer-to-Peer Networking, Peer-to-Peer Video Streaming, Network QOS, Network Performance Analysis, Network Security, Network Security Events Analysis and Correlation, Network Management, Cloud Computing and Networking, Software Defined Networks.
\end{minipage}%

\hfill \break

\noindent\begin{minipage}{0.1\textwidth}
\includegraphics[width=1.15in,height=1.15in,clip,keepaspectratio]{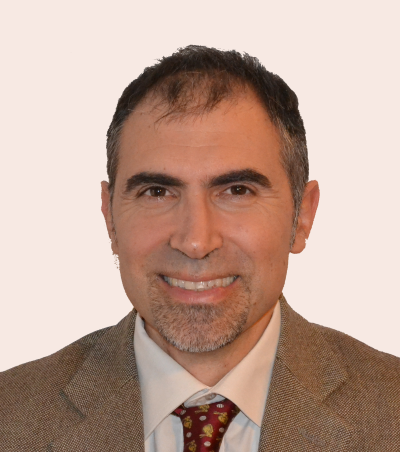}
\end{minipage}%
\hfill%
\begin{minipage}{0.82\textwidth}
\textbf{Stefano Salsano} received his PhD from University of Rome “La Sapienza” in 1998. He is Associate Professor at the University of Rome Tor Vergata. He participated in 15 research projects funded by the EU, being project coordinator in one of them and technical coordinator in two of them. He has been the principal investigator in several research and technology transfer contracts funded by industries. His current research interests include Software Defined Networking, Network Virtualization, Cybersecurity, Information-Centric Networking. He is a co-author of an IETF RFC and of more than 140 peer-reviewed papers and book chapters.
\end{minipage}%

\hfill \break

\noindent\begin{minipage}{0.1\textwidth}
\includegraphics[width=1.15in,height=1.15in,clip,keepaspectratio]{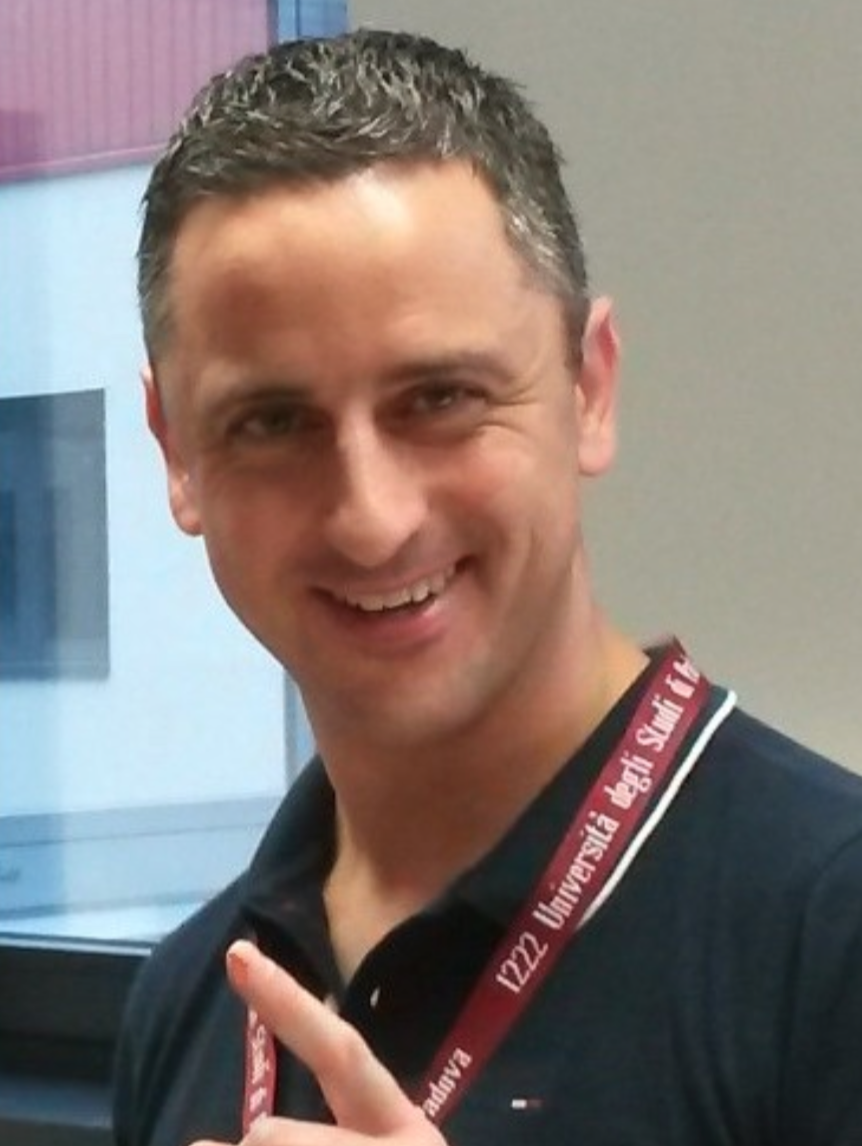}
\end{minipage}%
\hfill%
\begin{minipage}{0.82\textwidth}
\textbf{Mauro Conti} is a Professor at the University of Padua, Italy. His main research interest is in the area of security and privacy. In this area, he published more than 200 papers in topmost international peer-reviewed journals and conference. He is Associate Editor for several journals, including IEEE Communications Surveys \& Tutorials and IEEE Transactions on Information Forensics and Security. He was Program Chair for TRUST 2015, ICISS 2016, WiSec 2017, and General Chair for SecureComm 2012 and ACM SACMAT 2013. He is Senior Member of the IEEE.
\end{minipage}%

\hfill \break

\noindent\begin{minipage}{0.1\textwidth}
\includegraphics[width=1.15in,height=1.15in,clip,keepaspectratio]{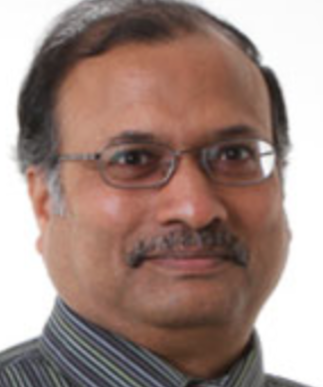} 
\end{minipage}%
\hfill%
\begin{minipage}{0.82\textwidth}
\textbf{Mukesh Singhal} is a Chancellor's professor and the chairman in the electrical engineering and computer science at the University of California, Merced. From 2001 to 2012, he was a professor and Gartner Group endowed chair in Network Engineering in the Department of Computer Science, University of Kentucky. His current research interests include distributed and cloud computing, cyber-security, and computer networks. He received 2003 IEEE Technical Achievement Award. He has published over 260 refereed articles in these areas. He has coauthored four books, including Advanced Concepts in Operating Systems, McGraw-Hill, New York, 1994 and Distributed Computing Systems. He has served in the editorial board of IEEE Transactions on Dependable and Secure Computing, IEEE Transactions on Parallel and Distributed Systems, IEEE Transactions on Data and Knowledge Engineering, and IEEE Transactions on Computers. He is a fellow of the IEEE.
\end{minipage}%

\end{document}